# A Survey of Machine and Deep Learning Methods for Internet of Things (IoT) Security

Mohammed Ali Al-Garadi, Amr Mohamed, Abdulla Al-Ali, Xiaojiang Du, Mohsen Guizani

*Abstract*—The Internet of Things (IoT) integrates billions of smart devices that can communicate with one another with minimal human intervention. It is one of the fastest developing fields in the history of computing, with an estimated 50 billion devices by the end of 2020. On the one hand, IoT technologies play a crucial role in enhancing several real-life smart applications that can improve life quality. On the other hand, the crosscutting nature of IoT systems and the multidisciplinary components involved in the deployment of such systems have introduced new security challenges. Implementing security measures, such as encryption, authentication, access control, network security and application security, for IoT devices and their inherent vulnerabilities is ineffective. Therefore, existing security methods should be enhanced to secure the IoT ecosystem effectively. Machine learning and deep learning (ML/DL) have advanced considerably over the last few years, and machine intelligence has transitioned from laboratory curiosity to practical machinery in several important applications. The ability to monitor IoT devices intelligently provides a significant solution to new or zero-day attacks. ML/DL are powerful methods of data exploration for learning about 'normal' and 'abnormal' behaviour according to how IoT components and devices perform within the IoT environment. Consequently, ML/DL methods are important in transforming the security of IoT systems from merely facilitating secure communication between devices to security-based intelligence systems. The goal of this work is to provide a comprehensive survey of ML methods and recent advances in DL methods that can be used to develop enhanced security methods for IoT systems. IoT security threats that are related to inherent or newly introduced threats are presented, and various potential IoT system attack surfaces and the possible threats related to each surface are discussed. We then thoroughly review ML/DL methods for IoT security and present the opportunities, advantages and shortcomings of each method. We discuss the opportunities and challenges involved in applying ML/DL to IoT security. These opportunities and challenges can serve as potential future research directions.

## I. INTRODUCTION

THE recent progress in communication technologies, such as the Internet of Things (IoT), has remarkably transcended the traditional sensing of surrounding environments. IoT technologies can enable modernisations that improve life quality [1] and have the capability to collect, quantify and understand the surrounding environments. This situation simplifies the new communication forms among things and humans and thus enables the realisation of smart cities [2]. IoT is one of the fastest emerging fields in the history of computing, with an estimated 50 billion devices by the end of 2020 [3, 4]. On the one hand, IoT technologies play a crucial role in enhancing real-life smart applications, such as smart healthcare, smart homes, smart transportation and smart education. On the other hand, the crosscutting and large-scale nature of IoT systems with various components involved in the deployment of such systems have introduced new security challenges.

IoT systems are complex and contain integrative arrangements. Therefore, maintaining the security requirement in a wide-scale attack surface of the IoT system is challenging. Solutions must include holistic considerations to satisfy the security requirement. However, IoT devices mostly work in an unattended environment. Consequently, an intruder may physically access these devices. IoT devices are connected normally over wireless networks where an intruder may access private information from a communication channel by eavesdropping. IoT devices cannot support complex security structures given their limited computation and power resources [5]. Complex security structures of the IoT are due to not only limited computation, communication and power resources but also trustworthy interaction with a physical domain, particularly the behaviour of a physical environment in unanticipated and unpredictable modes, because the IoT system is also part of a cyber-physical system; autonomously, IoT systems must constantly adapt and survive in a precise and predictable manner with safety as a key priority, particularly in settings where threatening conditions, such as in health systems, might occur [6]. Moreover, new attack surfaces are introduced by the IoT environment. Such attack surfaces are caused by the interdependent and interconnected environments of the IoT. Consequently, the security is at higher risk in IoT systems than in other computing systems, and the traditional solution may be ineffective for such systems [7].

A critical consequence of the extensive application of IoT is that IoT deployment becomes an interconnected task. For example, IoT systems should simultaneously consider energy efficiency, security, big IoT data analytics methods and interoperability with software applications [4] during the deployment stage. One aspect cannot be ignored when considering advances in another [4]. This integration provides

Mohammed Ali Al-garadi, Amr Mohamed, Abdulla Al-Ali, are with Department of Computer Science and Engineering, Qatar University, 2713, Doha, Qatar. E-mails: {mohammed.g, abdulla.alali, amrm}@qu.edu.qa Xiaojiang Du is with Department of Computer and Information Sciences, Temple University, Philadelphia. E-mail: xjdu@temple.edu. Mohsen Guizani is with Department of Electrical and Computer Engineering, University of Idaho, Moscow, Idaho, USA. E-mail: mguizani@ieee.org.

a new opportunity for researchers from interdisciplinary fields to investigate current challenges in IoT systems from different perspectives. However, this integration also introduces new security challenges due to the distribution nature of IoT devices, which provide a large and vulnerable surface. This characteristic of IoT devices presents many security issues. Moreover, the IoT platform generates a large volume of valuable data. If these data are not transmitted and analysed securely, then a critical privacy breach may occur.

IoT systems are accessible worldwide, consist mainly of constrained resources and constructed by lossy links [8]. Therefore, crucial modifications of existing security concepts for information and wireless networks should be implemented to provide effective IoT security methods. Applying existing defence mechanisms, such as encryption, authentication, access control, network security and application security, is challenging and insufficient for mega systems with many connected devices, with each part of the system having inherent vulnerabilities. For example, 'Mirai' is an exceptional type of botnets that has recently caused large-scale DDoS attacks by exploiting IoT devices[7, 9]. Existing security mechanisms should be enhanced to fit the IoT ecosystem [7]. However, the implementation of security mechanisms against a specified security threat is quickly conquered by new types of attacks created by attackers to circumvent existing solutions. For example, amplified DDoS attacks utilise spoofed source IP addresses for the attack location to be untraceable by defenders. Consequently, attacks that are more complex and more destructive than Mirai can be expected because of the vulnerabilities of IoT systems. Moreover, understanding which methods are suitable for protecting IoT systems is a challenge because of the extensive variety of IoT applications and scenarios [7]. Therefore, developing effective IoT security methods should be a research priority [7, 9].

As shown in Figure 1, having the capability to monitor IoT devices can intelligently provide a solution to new or zero-day attacks. Machine learning and deep learning (ML/DL) are powerful methods of data exploration to learn about 'normal' and 'abnormal' behaviour according to how IoT components and devices interact with one another within the IoT environment. The input data of each part of the IoT system can be collected and investigated to determine normal patterns of interaction, thereby identifying malicious behaviour at early stages. Moreover, ML/DL methods could be important in predicting new attacks, which are often mutations of previous attacks, because they can intelligently predict future unknown attacks by learning from existing examples. Consequently, IoT systems must have a transition from merely facilitating secure communication amongst devices to security-based intelligence enabled by DL/ML methods for effective and secure systems.

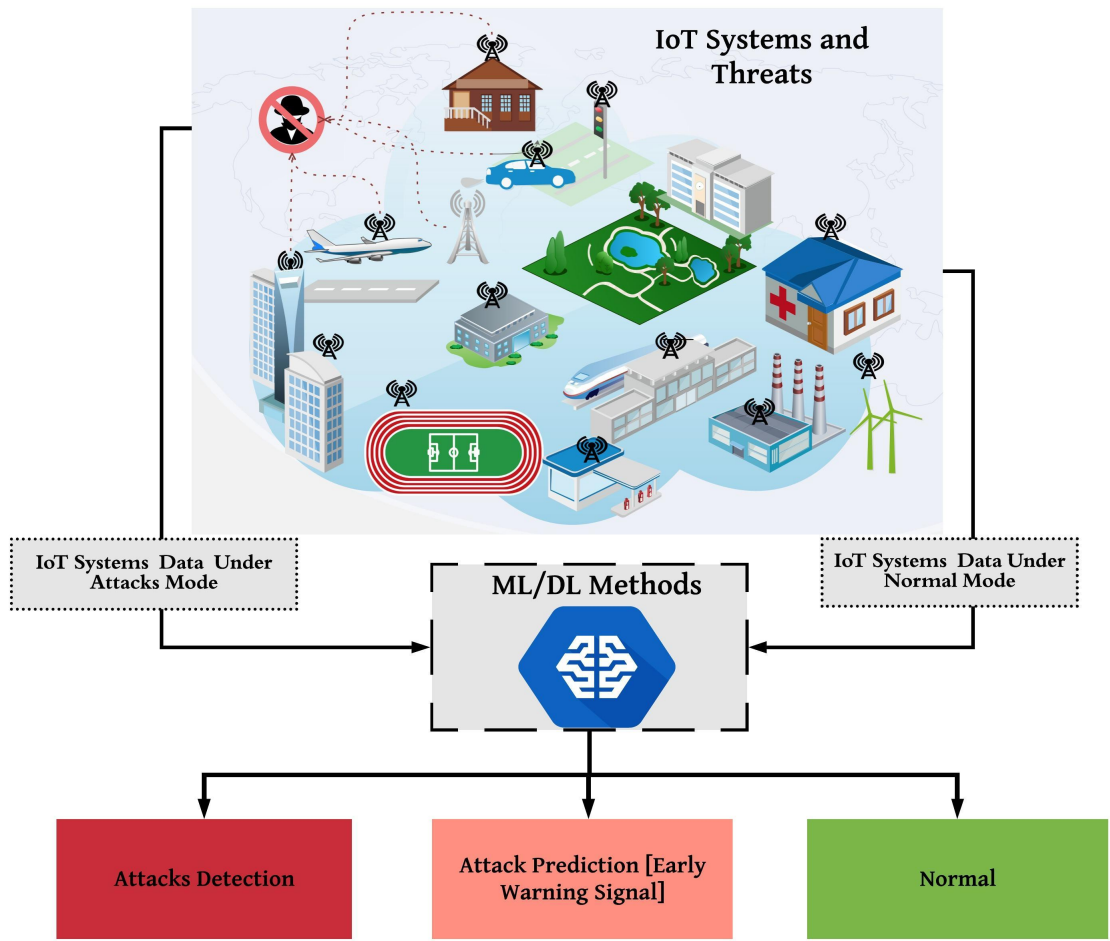

*Figure 1 Illustration of the potential role of ML/DL in IoT security*

Although essentially DL is a branch of ML, this paper discusses them in two separate sections to provide the readers with in-depth review, inclusive comparisons, and potential applications of both traditional ML and DL methods for IoT security. The main differences between traditional ML and DL methods has been discussed in previous literature [10, 11]. Similarly, in this paper ML refers to traditional ML methods that require engineered features, while DL methods refer to recent advances in learning methods that utilise several non-linear processing layers for discriminative or generative feature abstraction and transformation for pattern analysis [12].

Figure 2 shows a thematic taxonomy of ML/DL for IoT security. The remaining parts of the paper adopt the classification presented on the thematic taxonomy. The present survey comprehensively reviews ML/DL algorithms for IoT security that can provide researchers and developers a manual guide to developing an effective and end-to-end security solution based on intelligence. This survey also aims to highlight the list of challenges of using ML/DL to secure IoT systems. Section II provides an overview of general IoT systems, but the purpose of such an overview is to summarise the method used by the IoT model and its characteristics for increasing security risk. The summary points are provided at the end of Section II. Section III presents the IoT security properties and threats and discusses the potential vulnerabilities and attack surfaces of IoT systems (IoT attack surfaces are categorised into physical device, network service, cloud service and web and application interfaces). Moreover, we discuss a new attack surface caused by the IoT environment. In Section IV, we discuss the most promising ML and DL algorithms, their advantages, disadvantages and applications in the IoT security and then present the comparison and summary table for the reviewed ML/DL methods at the end of each section. Section V discusses and comprehensively compares the application of ML/DL methods in securing each IoT layer, and a summary table of the studies that used ML and DL for IoT security is presented. In this section, we also present the enabling technology of ML/DL deployment for IoT security. In Section VI, the issues, challenges and future directions in using ML/DL for effectively securing IoT systems are presented and classified; the challenges are related to IoT data issues, learning strategies, operations under the interdependent, interconnected and interactive environments, possible misuse of ML and DL algorithms by attackers, inherent privacy and security issues of ML and DL and inherent properties of an IoT device. These challenges prevent the implementations of effective ML/DL methods for IoT system security (i.e. computational complexity or security vs. other trade-offs) and are presented as future directions. Furthermore, we present other future directions, such as integrating ML/DL with other technology (e.g. edge computing and blockchain) to provide reliable and effective IoT security methods. Section VII presents the conclusions drawn from this survey.

The key contributions of this survey are listed as follows:
- *Comprehensive discussion on the potential vulnerabilities and attack surfaces of IoT systems*: We discuss various threats and attack surfaces in IoT systems. The attack surfaces are categorised into physical device, network service, cloud service and web and application interfaces, with several examples of security threat and potential vulnerabilities for each attack surfaces. We also discuss a new attack surface caused by the interdependent, interconnected and interactive environments of IoT systems.
- *In-depth review of the ML and recent advances in DL methods for IoT security*: The most promising ML and DL algorithms for securing IoT systems are reviewed, and their advantages, disadvantages and applications in IoT security are discussed. Furthermore, comparisons and summary tables for ML and DL methods are presented to provide learned lessons.
- *Application of ML/DL for each IoT layers:* The application of ML/DL for securing perception, network and application layers is reviewed. The works reviewed are compared on the basis of the type of learning method used, the type of attack surfaces secured and the type of threats detected. The enabling technologies of ML/DL deployment for IoT security are discussed.
- *Challenges and future directions:* Several potential research challenges and future directions of ML/DL for IoT security are presented.

The following subsection discusses related works to highlight the major differences of this survey from the previous survey on IoT security.

*A. Related Work*

Several researchers have conducted surveys on the IoT security to provide a practical guide for existing security vulnerabilities of IoT systems and a roadmap for future works. However, most of the existing surveys on IoT security have not particularly focused on the ML/DL applications for IoT security. For example, Surveys [13-19] reviewed extant research and classified the challenges in encryption, authentication, access control, network security and application security in IoT systems. Granjal, Monteiro and Silva [20] emphasised the IoT communication security after reviewing issues and solutions for the security of IoT communication systems. Zarpelão et al. [21] conducted a survey on intrusion detection for IoT systems. Weber [22] focused on legal issues and regulatory approaches to determine whether IoT frameworks satisfy the privacy and security requirements. Roman, Zhou and Lopez [23] discussed security and privacy in the distributed IoT context. These researchers also enumerated several challenges that must be addressed and the advantages of the distributed IoT approach in terms of security and privacy concerns. Survey [24] reviewed evolving vulnerabilities and threats in IoT systems, such as ransomware attacks and security concerns. Xiao et al. [25] briefly considered the ML methods for protecting data privacy and security in the IoT context. Their study also indicated three challenges in future directions of ML implementation in IoT systems (i.e. computation and communication overhead, backup security solutions and partial state observation). Other survey papers such as [26, 27] focused on the uses of data mining and machine learning methods for cybersecurity to support intrusion detection. The surveys mainly discussed the security of the cyber domain using data mining and machine learning methods and mainly reviewed misuse and anomaly detections in cyberspace [26, 27].

However, in contrast to other surveys, our survey presents a comprehensive review of cutting-edge machine and recent advances in deep learning methods from the perspective of IoT security. This survey identifies and compares the opportunities, advantages and shortcomings of various ML/DL methods for IoT security. We discuss several challenges and future directions and present the identified challenges and future directions on the basis of reviewing the potential ML/DL applications in the IoT security context, thereby providing a useful manual for researchers to transform the IoT system security from merely enabling a secure communication among IoT components to end-to-end IoT security-based intelligent approaches.

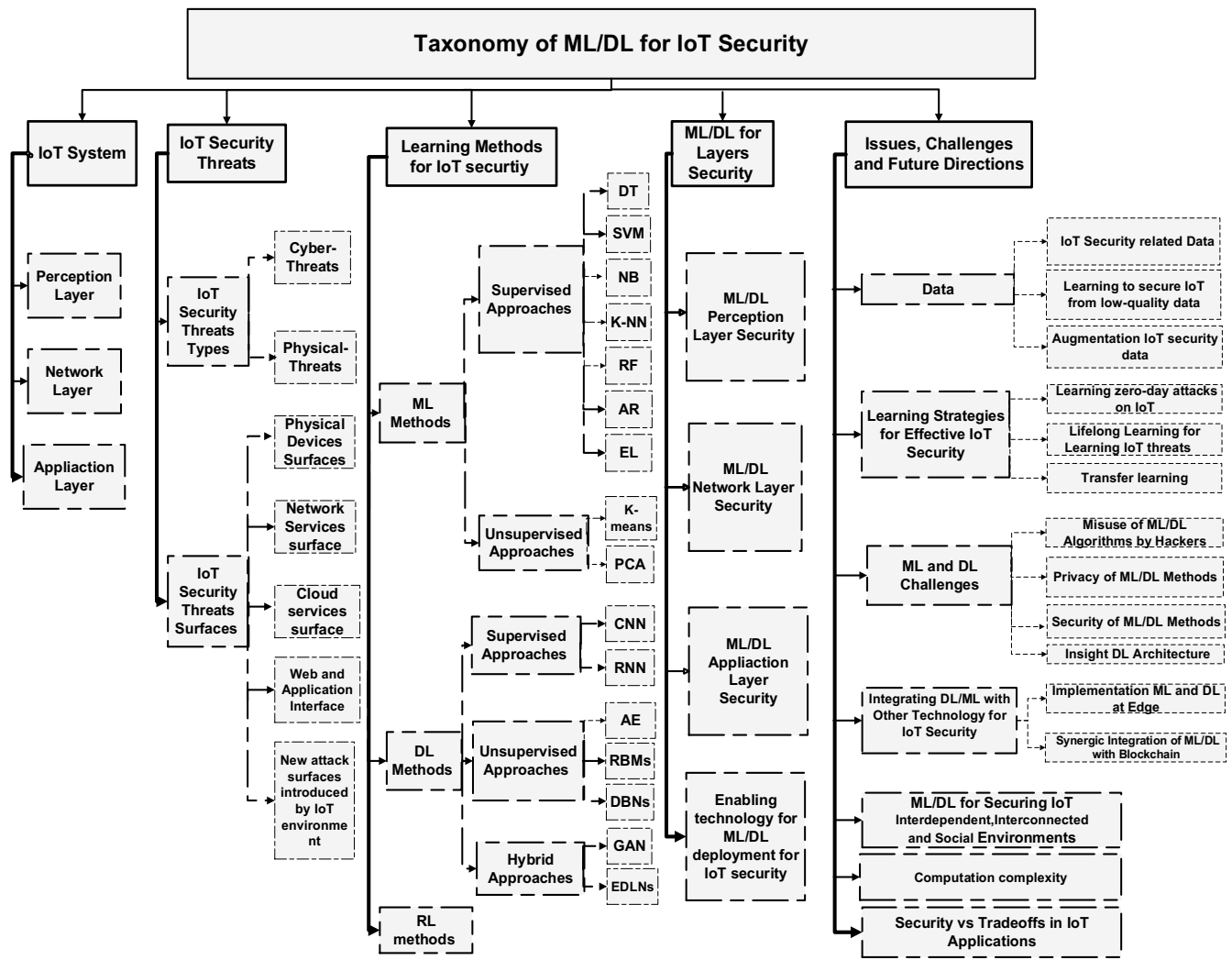

*Figure 2 Thematic Taxonomy of ML/DL for IoT Security*

## II. OVERVIEW OF THE IOT SYSTEM

This section provides an overview of the general IoT systems. However, the objective of this section is to highlight the characteristics of IoT systems that may increase security risk. The summary points are provided at the end of this section.

IoT converts a physical object from a conventional object to a smart object by utilising technologies, such as communication technologies, Internet protocols and applications, sensor networks and ubiquitous and pervasive computing [28]. The implementation of a flawless IoT system is crucial in the academe and industry due to the wide range of applications that can enable the execution of smart city concepts through billions of connected smart devices [29]. The IoT model can be defined as the interconnection of massive heterogeneous devices and systems in diverse communication patterns, such as thing-to-thing human-to-human or human-to-thing [15, 28]. The IoT architecture consists of physical objects that are integrated into a communication network and supported by computational equipment with the aim of delivering smart services to users. The IoT architecture generally has three layers, namely, application, network and perception [16]. This architecture can be further taxonomized for simplicity and improved analysis, as shown in Figure 3. Each level is described in the following subsections.

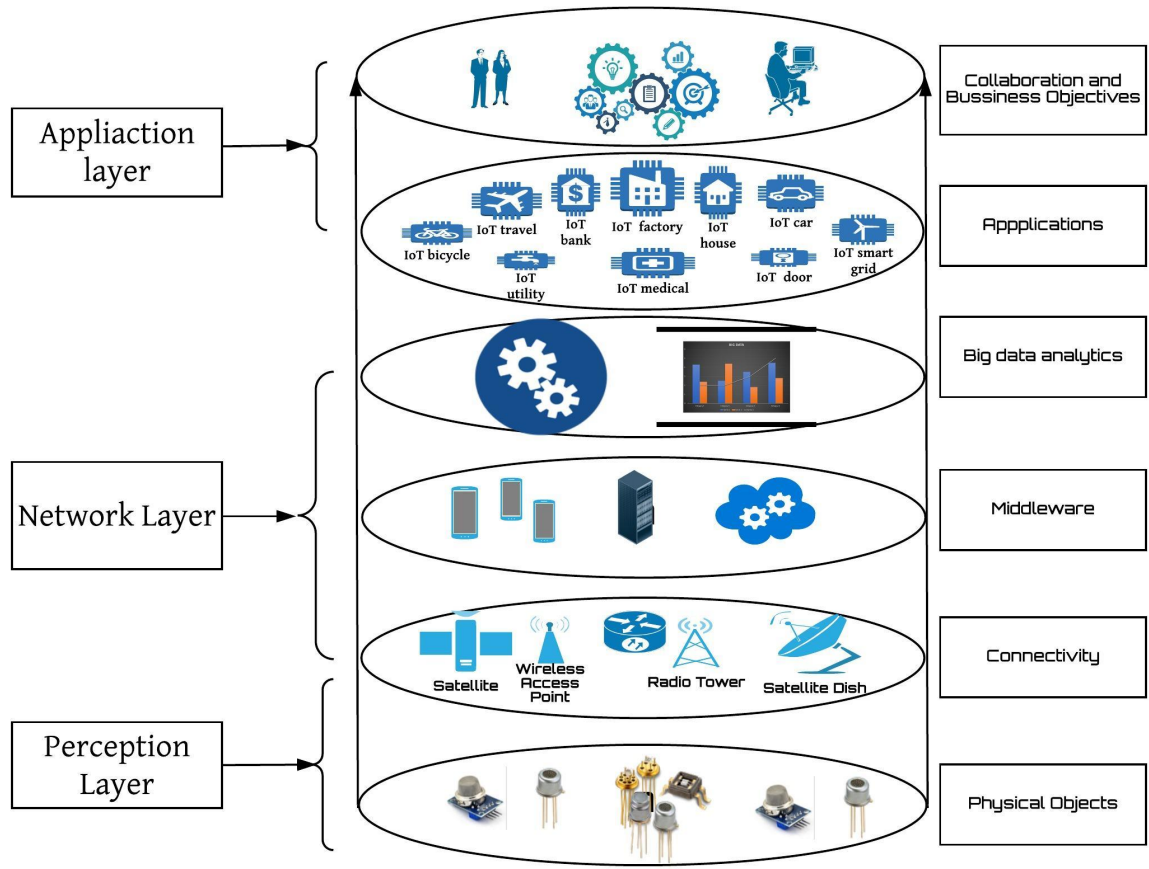

*Figure 3 IoT architecture*

### A. Physical objects

The physical object level involves IoT physical sensors. The main function of physical objects is to sense, collect and possibly process information. This level adopts sensors and actuators, such as temperature, humidity, motion and acceleration sensors, to implement diverse sensing functionalities. The plug-and-play mechanism must be applicable at this level to configure heterogeneous sensors [28, 30, 31]. IoT sensors are resource-constrained devices because they have limited battery capacity and computation capability. Understanding the sensor data delivered by these objects is a key step in achieving a context-aware IoT system [32, 33]. A large part of the big data of IoT is generated at this level. The increase in the number of IoT devices and the extensive increment in data volume indicate a positive correlation between the growth of big data and the growth of IoT devices. Effective analysis of the big data of IoT can result in improved decision making for a highly secure IoT implementation.

### B. Connectivity

One of the main objectives of the IoT platform is to connect heterogeneous sensors cooperatively and subsequently provide smart services [28]. The sensors implemented in the IoT platform are resource-constrained because they are powered by batteries and have a limited computation and storage capability [33, 34]. Therefore, IoT sensors must work with low-power resources under a lossy and noisy communication environment [28]. The following connectivity challenges are encountered in the deployment of IoT devices.

- The first one is providing unique IPs to billions of devices connected to the Internet. This challenge can be mitigated by incorporating 6LoWPAN that uses IPv6.
- The second challenge is developing low-power communication for transmitting data generated by sensors.
- The third challenge is implementing effective routing protocols that consider the limited memory of sensors and support the flexibility and mobility of smart objects.

The recent communication technologies employed in IoT are 6LoWPAN, Bluetooth, IEEE 802.15.4, WiFi, ultra-wide bandwidth, RFID and near-field communication (NFC) [28].

### C. Middleware

A middleware aims to effectively represent the complexities of a system or hardware, thus allowing developers to focus only on the issue to be solved without interruption at the system or hardware level [35, 36]. These complexities are commonly related to communication and computational issues. A middleware offers a software level amongst applications, the operating system and the network communication levels; it enables cooperative processing. From the computational perspective, a middleware offers a level between an application and the system software [33, 35, 37]. Its main functions can be summarised as follows. First, it enables cooperation between heterogeneous IoT objects so that the diverse categories of IoT can interact with one another effortlessly through middleware assistance [33, 35, 37]. One of key roles of middleware is to provide interoperability between the IoT devices. Second, a middleware must provide scalability amongst several devices that are likely to interact in the IoT realm. The future growth of IoT devices should be handled by the middleware by providing vital modifications when the organisation scales [33]. The third function is device discovery [33] and context awareness, which should be provided by a middleware to support the objects' awareness of all other surrounding IoT objects. A middleware should provide context-aware computing to understand sensor data. Sensor data can be utilised to obtain the context, and the obtained context can be used to provide smart services to users [38]. The last function is to provide security and privacy to IoT devices because the data collected by IoT devices are generally related to humans or an industry. Security and privacy concerns must be addressed in such circumstances. A middleware must construct mechanisms to provide a secure IoT system [33].

*D. Big Data Analytics*

The huge amounts of data produced or captured by IoT are extremely valuable. ML can play an analytical role in building intelligent IoT systems to deliver smart services in the IoT realm [39]. Big data are created [40] by several physical objects that are used in various IoT applications. However, physical devices produce volumes of data that should be analysed in real time to acquire useful knowledge. To obtain insights from these data, researchers [39-44] have discussed different methods of integrating big data analytical methods with IoT design. Unlike traditional analytical methods, ML and DL can effectively derive unobserved insights from big data and convert big data into useful data with minimal human assistance [40]. Analytical methods can be categorised into three: descriptive, predictive and perspective analytics [40]. Descriptive analytics is used for analysing data to describe current or past events. Predictive analytics is used for analysing data to predict the future based on the patterns that occur in current events. Prescriptive analytics is used for analysing data to make decisions by examining various real scenarios and providing a set of recommendations to decision makers. The big data related to the behaviour of IoT systems are vital in building ML/DL to secure IoT systems.

*E. Applications*

IoT has several applications. The commonly known applications include smart healthcare, smart transportation, smart grid and smart building. These applications are briefly discussed in the following subsections.

*1) Smart healthcare*

IoT devices have become popular in health applications in recent years. The IoT system is rapidly becoming a key instrument in healthcare [45]. IoT devices are used in healthcare sectors to closely observe and record patient conditions and send warnings to the concerned healthcare system in critical circumstances to provide a rapid and timely treatment to patients. Internet of medical things (IoMT) devices have been adopted in approximately 60% of the healthcare sector [46]. IoMT is believed to have a significant role in transforming the healthcare field by empowering the evolution from disorganised healthcare to synchronised healthcare. In 2015, 30.3% of 4.5 billion IoT devices are IoMT devices. This number is estimated to rise to 20–30 billion IoMT devices by 2020.

However, in contrast to other applications, the IoT in healthcare systems must be secured whilst providing flexible access to devices to save lives in emergency cases. For example [47], an individual who has an implanted IoT-based medical device has experienced an emergency situation where he/she suddenly must be admitted into a hospital rather than only regularly visiting. In this case, the staff at the new hospital must be able to access the implanted IoT-based medical device easily. Therefore, a complex security requirement may be not acceptable, and the security method must consider and balance between security and flexible access during emergency situations.

Moreover, IoT sensors are widely used to monitor daily health-related activities. A smartphone is usually employed to monitor health-related activities, such as daily activity (number of steps, walking and running distance and cycling distance), and sleep analysis. IoT has prodigious opportunities to potentially advance healthcare systems and a wide range of applications [48]. The recent development in traditional medical devices towards interactive environment medical devices can be further advanced by the IoT system by connecting implanted, wearable and environmental sensors collaboratively within the IoT system to monitor users' health effectively and ensure real-time health support [45]. However, securing IoT systems remains a critical issue [49, 50], and further investigation is required to securely implement IoT devices in healthcare.

*2) Smart transportation*

Smart or intelligent transport systems have become attainable with the help of IoT systems. The main objective of smart transport is to manage daily traffic in cities intelligently by analysing data from well-connected sensors located in different places and implementing data fusion (data from CCTV, mobile devices, GPS, accelerometers, gyroscope-based applications and weather sensors). The data are then explored and integrated to provide smart choices to users [51]. Moreover, the data analytics of smart transport can implicitly enhance shipment schedules, advance road safety and improve delivery time [40].

*3) Smart governance*

IoT can facilitate smart governance. Integrating data from different governmental sectors can provide authorities with abundant information from a wide range of sensor data (from weather-related data to security-related data). The huge amount of data generated by IoT sensors can overcome the limitations of conventional monitoring systems in an exceptional manner, thereby presenting a knowledge-based system from information fusion sources that compiles and correlates data from different sectors to deliver an optimal decision considering multiple perspectives.

*4) Smart agriculture*

IoT systems can be applied to improve the agriculture sector. IoT sensors can be implemented to enable real-time monitoring of the agriculture sector. IoT sensors can collect useful data on humidity level, temperature level, weather conditions and moisture level. The collected data can then be analysed to provide important real-time mechanisms, such as automatic irrigation, water quality monitoring, soil constituent monitoring and disease and pest monitoring [52].

*5) Smart grid*

The latest development in power grids was achieved by using the IoT platform to construct a smart grid in which the electricity between suppliers and consumers is handled smartly to improve efficiency, safety and real-time monitoring[53-55]. The IoT platform plays a significant role in effective grid management. Applying IoT technology in a smart grid can help prevent disasters, decrease power transmission to enhance the reliability of power transmission and minimise economic losses [56]. Moreover, analysing the data generated by IoT sensors can help decision makers select a suitable electricity supply level to deliver to customers.

*6) Smart homes*

IoT components are used to realise smart homes. Home IoT-based machines and systems (e.g. fridge, TV, doors, air conditioner, heating systems and so on) are now easy to observe and control remotely [28, 57]. A smart home system can understand and respond to surrounding changes, such as automatically switching on air conditioners based on weather predictions and opening the door based on face recognition. Intelligent homes should consistently collaborate with their internal and external environments [58]. The internal environment involves all home IoT devices that are managed internally, and the external environment involves objects that are not managed by the smart home but play important roles in the construction of the smart home, such as smart grids [28, 58].

*7) Smart supply chain*

An important application of IoT technology in real life is the development of easier and more flexible business processes than before. The development in IoT-embedded sensors, such as RFID and NFCE, enables the interaction between IoT sensors embedded on the products and business supervisors. Therefore, these goods can be tracked throughout production and transportation processes until they reach the consumer. The monitoring process and the data generated through this process are crucial in making appropriate decisions, which can in turn improve machine uptime and the service provided to customers [55].

*F. Collaboration and Business Objective*

At this level, IoT service is delivered to users, and the data captured and analysed at the lower levels are integrated into the business objective. This level mainly involves human interaction with all of the levels of the IoT model. The aim at this level is to effectively utilise the data captured, transmitted and analysed at different levels of the IoT model to improve social and economic growth. The analysis of big data generated by IoT devices can be incorporated into the business objective at this stage to identify factors that can improve the business outcome and create optimal strategic business plans.

From the above discussions of the IoT system, we can conclude that the nature of IoT systems can increase security risks because of the following reasons:

- By nature, the IoT is a multipart model with various applications with diverse requirements. This nature demonstrates the huge complexity of such systems through extensive IoT applications, from smart home, smart car to smart healthcare. Such drivers and various IoT applications can be a challenge whilst developing an effective security scheme, in which, the effective security method proposed for a specific application or requirement may be unsuitable for other applications with different requirements.
- IoT systems are vastly heterogeneous in protocols, platforms and devices that are accessible worldwide, consist mainly of constrained resources, constructed by lossy links [7, 8] and lack standardization. Such features of IoT systems become bottlenecks that prevent the development of effective and generalised security schemes for such systems.
- IoT devices can be designed to autonomously adapt to the surrounding environment. Consequently, IoT devices can be controlled by other devices [7]. In such cases, an effective IoT security must not only be proposed to secure each device independently but also to provide an end-to-end security solution.
- The data generated by the IoT is valuable, and such data can be analysed to understand the behaviour of individuals and their daily activities. Therefore, such information can be used by policymakers to adjust their products smartly to satisfy the preferences and requirements of individuals. However, this result can turn IoT devices into eavesdropping devices that capture user information including biometric data, such as voices, faces and fingerprints, that can aid in IoT device intrusion.
- Physical attacks can increase by implementing IoT systems because most of the physical things of IoT (e.g. sensors) may be ubiquitously and physically reachable [23, 59]. Physical threats may also be caused by unintended damage from natural disasters, such as floods or earthquakes, or disasters caused by humans, such as wars [60, 61]. Therefore, an effective security solution must be context-aware by considering such characteristics of IoT systems.
- IoT systems do not have exact boundaries and are constantly adjusted whilst new devices are added or

due to user mobility. Such characteristics allow the IoT model to continually expand attack surfaces and introduce several vulnerabilities.

*Therefore*, methods that can comprehensively understand and gain knowledge on the behaviour of things and other IoT components within such large systems are required. However, ML/DL methods can predict the expected behaviour of a system by learning from previous experiences. Therefore, applying ML/DL methods can significantly advance the security methods by transforming the security of IoT systems from simply facilitating secure communication between devices to security-based intelligence systems.

### III. IoT Security Threats

IoT integrates the Internet with the physical world to provide an intelligent interaction between the physical world and its surroundings. Generally, IoT devices work in diverse surroundings to accomplish different goals. However, their operation must meet a comprehensive security requirement in cyber and physical states [60, 62]. IoT systems are complex and contain multidisciplinary arrangements. Therefore, maintaining the security requirement with the wide-scale attack surface of the IoT system is challenging. To satisfy the desired security requirement, the solution should include holistic considerations. However, IoT devices mostly work in an unattended environment. Consequently, an intruder may physically access these devices. IoT devices are normally connected over wireless networks where an intruder might expose private information from the communication channel by eavesdropping. IoT devices cannot support complex security structures because of their limited computation and power resources [5]. Therefore, securing the IoT system is a complex and challenging task. Given that the main objective of the IoT system is to be accessed by anyone, anywhere and anytime, attack vectors or surfaces also become accessible to attackers [23, 63]. Consequently, causing potential threats to become more probable. A threat is an act that can exploit security weaknesses in a system and exerts a negative impact on it [5]. Numerous threats, such as passive attacks (e.g. eavesdropping) and active threats (e.g. spoofing, Sybil, man-in-the-middle, malicious inputs and denial of service (DoS)), might affect the IoT system. Figure 4 shows the potential attacks that can affect the main security requirements (authentication, integrity non-repudiation, confidentiality availability and authorisation).

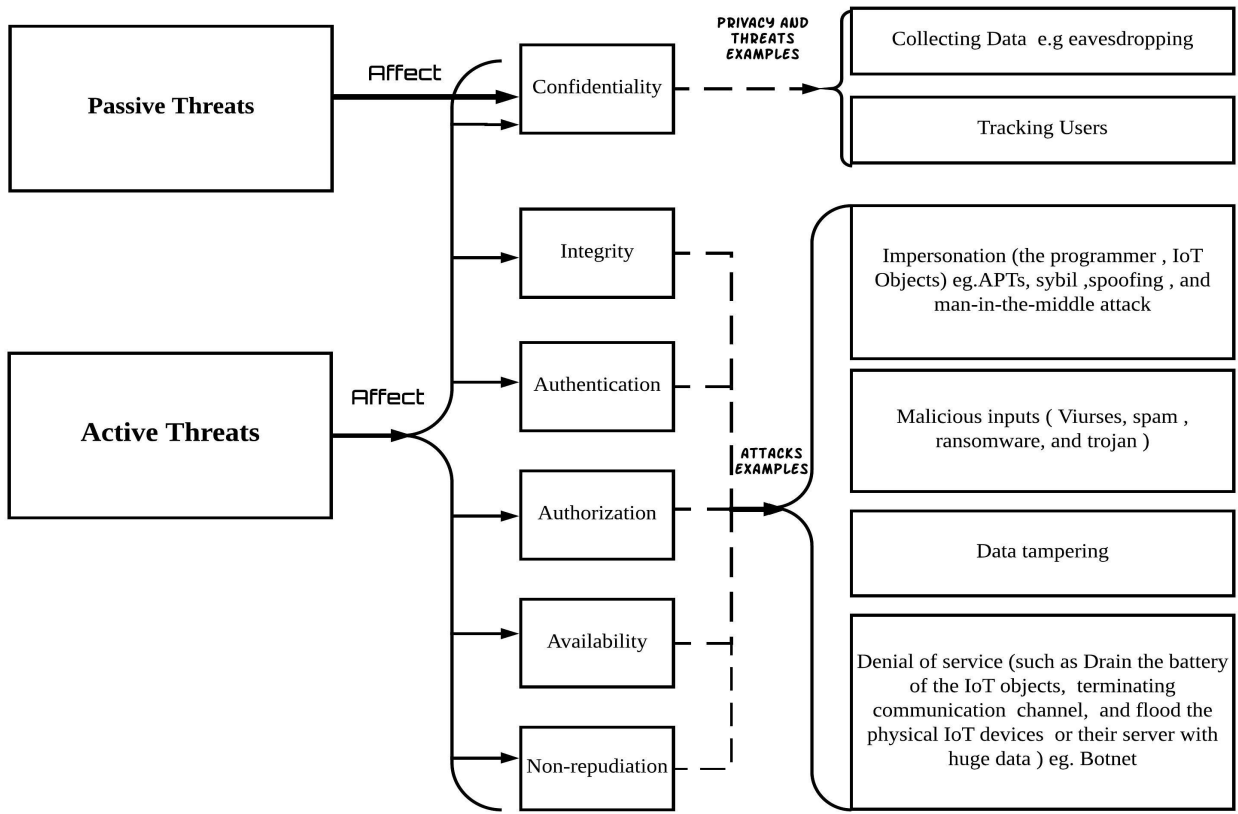

*Figure 4 Potential threats in the IoT system*

The following main security properties should be considered while developing an effective IoT security methods.

***Confidentiality***: Confidentiality is a vital security characteristic of IoT systems. IoT devices may store and transfer sensitive information that should not be revealed by unauthorised individuals. Medical (patient related), personal, industry and military data are highly confidential and must be secured against unauthorised access [5, 64]. However, in specific scenarios, such as IoT medical devices, although communications are encrypted and data are confidentially stored and transferred, attackers can still sense the existence of the physical device and can even track the holder. In such a situation, the location confidentiality of the holder is exposed and put at risk [47].

*Integrity*: Data from IoT devices are generally transferred through wireless communication and must be changed only by authorised entities. Integrity features are thus fundamental in ensuring an effective checking mechanism to detect any modification during communication over an insecure wireless network. Integrity features can secure the IoT system from malicious inputs that might be used to launch structured query language (SQL) injection attacks [65]. A deficiency in integrity inspection can allow for modification of the data stored on the memory of IoT devices, which can affect the main operational functions of the physical devices for a long time without being detected easily. IoT systems have various integrity requirements. For example, IoT implantable medical devices require effective integrity checking against random errors because they affect human lives directly. Loss, errors or modification of information in several circumstances can lead to loss of human lives [47, 66, 67].

*Authentication*: The identity of entities should be perfectly established prior to performing any other process. However, due to the nature of IoT systems, authentication requirements differ from system to system. For example, authentication should be robust in an IoT system where a service needs to offer robust security rather than high flexibility. Trade-offs are a major challenge in developing an effective authentication scheme. For example, the trade-off between security and safety in IoT medical devices is that both security and safety must be balanced when designing an authentication scheme. Similarly, the trade-off can be between an effective authentication scheme and battery-based devices or between privacy and security [68]. Therefore, an IoT system requires an effective authentication that can balance system constraints and provide robust security mechanisms [69].

*Authorisation*: Authorisation includes granting users access rights to an IoT system, such as a physical sensor device. The users may be machines, humans or services. For example, the data collected by sensors should only be delivered to and accessed by authorised users (authorised objects and service requesters) [15, 70]. In other words, an action must be performed only if the requester has satisfactory authorisation to command it. The main challenge in authorisation in IoT environments is how to grant access successfully in an environment where not only humans but also physical sensors (things) should be authorised to interact with the IoT system [14]. In addition, in handling huge amounts of data in such a heterogeneous environment, the data must be protected throughout the sensing and transmission process and should be made available only to authorised parties [71].

*Availability*: The services delivered by IoT systems must always be available to authorised entities. Availability is a fundamental feature of a successful deployment of IoT systems. However, IoT systems and devices can still be rendered unavailable by many threats, such as DoS or active jamming. Therefore, ensuring the continuous availability of IoT services to users is a critical property of IoT security.

*Non-repudiation*: The non-repudiation property is meant to provide access logs that serve as evidence in situations where users or objects cannot repudiate an action. Generally, non-repudiation is not considered a key security property for many IoT systems [5]. However, non-repudiation can be an important security property in specific contexts, such as payment systems where both parties cannot repudiate a payment transaction [5].

For an effective IoT security scheme, the security properties above should be considered. However, these properties can be exploited by several security threats, as shown in Figure 4. In the following subsection, we briefly discuss potential security threats to understand how different security properties should be maintained in a secure IoT environment.

*A. Threats in IoT*

Security threats can be categorised as cyber and physical. Cyber threats can be further classified as passive or active. The following subsection provides a brief discussion of these threats.

*1) Cyber Threats*

*Passive threats:* A passive threat is performed only by eavesdropping through communication channels or the network. By eavesdropping, an attacker can collect information from sensors, track the sensor holders, or both. Currently, collecting valuable personal information, particularly personal health data, has become rampant on the black market [62]. The value of personal health information on the black market is $50 compared with $1.50 for credit card information and $3 for a social security number [62]. Moreover, an attacker can eavesdrop on communication channels to track the location of the IoT device holder if its communication channel is within range [72, 73], thus causing a violation of privacy.

*Active threats*: In active threats, the attacker is not only skilful in eavesdropping on communication channels, but also in modifying IoT systems to change configurations, control communication, deny services and so on. Attacks may include a sequence of interventions, disruptions and modifications. For example, potential attacks on an IoT system (shown in Figure 4) may involve the following active attacks: impersonation (e.g. spoofing, Sybil and man-in-the-middle), malicious inputs, data tampering and DoS. An impersonation attack is intended to impersonate an IoT device or authorised users. If an attack path exists, active intruders can attempt to partially or fully impersonate an IoT entity [23]. Malicious input attacks are intended to insert malicious software into the targeted IoT system. This software will run a code injection attack. The injected malicious software has a dynamic nature, and new types of attacks are constantly introduced to violate IoT components remarkably because IoT systems have a naturally large, well-connected surface [24, 74]. Meanwhile, data tampering is the act of intentionally changing (deleting, changing, manipulating or editing) information via unauthorised operations. Data are commonly transmitted or stored. In both situations, data might be captured and tampered, which might affect the significant functions of IoT systems, such as changing the billing price in the case of an IoT-based smart grid [75]. Many types of DoS attacks can be utilised against IoT. These types range from conventional Internet DoS attacks that are established to deplete the resources of the service provider and network bandwidth to signal jamming that targets wireless communication. Distributed DoS (DDoS) is a severe DoS where several attacks are launched with different IPs, which makes discriminating it from normal traffic of normal devices challenging compared with the attack with a huge traffic form signal or limited number of devices that is easier to discriminate from normal traffic and devices. Although different forms of DoS attacks exist, they have a common aim: to interfere with the availability of IoT services [21]. IoT systems have billions of connected devices that can be exploited through destructive DDoS, such as Mirai. Mirai is an exceptional type of botnets that has recently caused large-scale DDoS attacks by using IoT devices [7, 9].

*2) Physical Threats*

Physical threats can be in terms of physical destruction. In these threats, the attacker generally does not have technical capabilities to conduct a cyber-attack. Therefore, the attacker can only affect the reachable physical objects and other components of IoT that lead to terminating the service. By adopting IoT systems, these types of attacks may become wide-scale because most of the physical objects of IoT (sensors and cameras) are expected to be everywhere and physically accessible [23, 59]. Physical threats may also be caused by unintended damage from natural disasters, such as floods or earthquakes, or disasters caused by humans, such as wars [60, 61].

## B. Attack Surfaces

In this section, we discuss possible IoT system attack surfaces and the potential threats related to each surface. As shown in Figure 5, IoT attack surfaces can be categorised into physical device, network service, cloud service, web and application interface. The new IoT environment introduces threat surfaces.

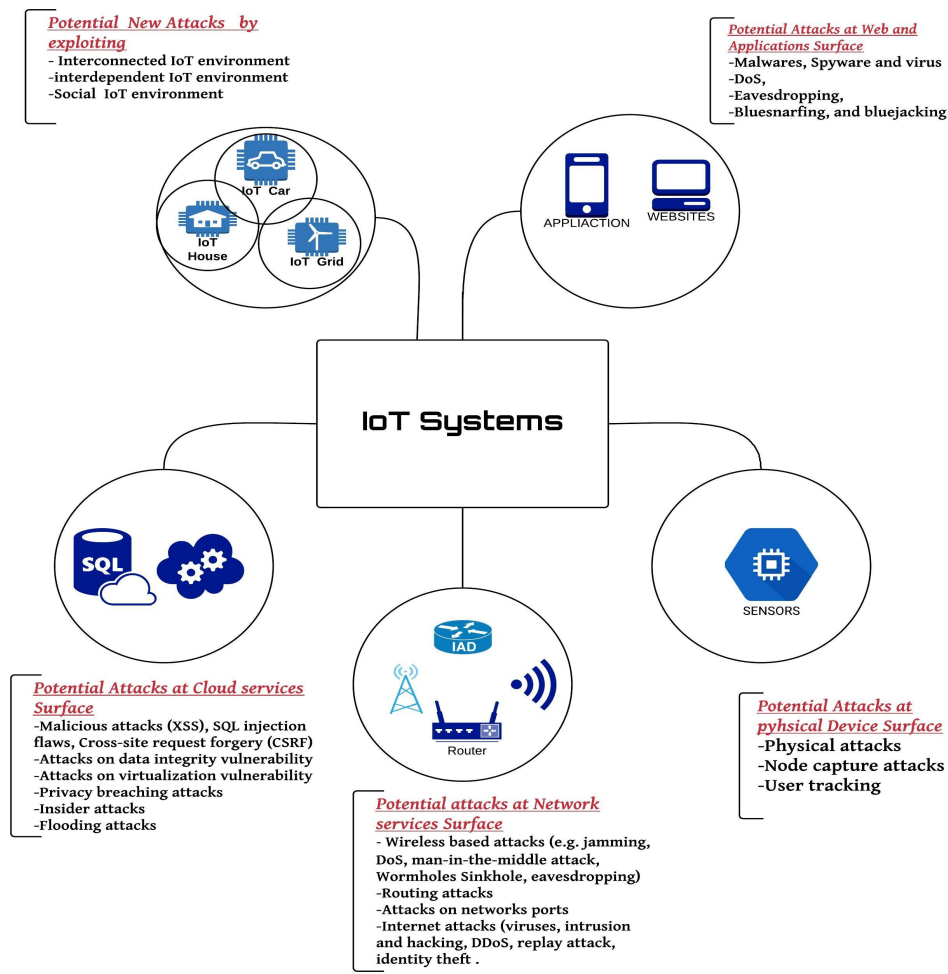

*Figure 5 IoT attack surfaces*

*1) Physical Device Surface*

Physical devices, such as RFID, are a main part of IoT systems. In an embedded communication system, RFID plays a significant role in implementing microprocessors for wireless communication [76]. The key characteristic of RFID tags is automatic identification through a unique identifier that involves fast information transmission between tags (RFID is tagged to an object that can be anything, from human to animal) and readers [77]. The main function of RFID technology is to supervise the process of objects in real time; this bridges the interaction between virtual and actual worlds. Consequently, these tiny physical devices can be expended in an exceptionally wide range of applications [76]. However, most physical devices suffer from many security-related issues. Another unit of physical device surface is the sensor node. Sensor nodes mainly consist of sensors used for sensing and actuators used for actuating devices in accordance with specified instructions. Sensors nodes commonly have high latency.

Most physical devices are resource-constrained and contain valuable information, which makes them a potential surface for attackers; for example, they can be exploited to track their holders, flooding them with many access requests that cause DoS or other attacks, such as eavesdropping, spoofing and counterfeiting [78, 79]. Moreover, this surface is highly vulnerable to physical threats because it is the most physically accessible surface for an attacker.

*2) Network Service Surface*

The IoT system contains physical objects (sensors and actuators) that are connected through wired and wireless technologies. Sensor networks (SNs) are significant resources for realising IoT systems. SNs can be constructed without an IoT system. However, an IoT system cannot be constructed without SNs [38, 80]. An IoT system consists of SNs and a wired network, thus creating a large-scale network surface. Such a wide network surface can be potentially vulnerable. Moreover, IoT systems face new security threats that are inherited from wired and wireless SNs. These new threats are introduced when traditional networks are directly integrated into IoT networks. The direct integration of a wireless SN into an IoT network poses several issues because traditional networks are no longer secure within IoT environments; for example, the resilience of WSNs (the sensors within a WSN openly provide its information to external parties) makes this network completely vulnerable to attacks in IoT environments [38, 80].

Other threats can be designed by attackers to target the routing protocol that may lead to network failure. Accordingly, designing a secure routing protocol is important to IoT system security [78, 81]. Attacks can also be launched at a port by searching and examining open ports. Detection of open ports can encourage attackers to launch an attack on the services operating on these open ports. Such an attack can extract detailed information about the network, such as IP address, MAC address, router and gateway [9, 82].

IoT has expanded network connectivity, mobility and collaboration between users. Such features increase network service surfaces, leading to frequent security risks and attacks, such as hacking, interruption, acknowledgement spoofing, DoS, man-in-the-middle attack protocol tunnelling and interception [83]. Furthermore, the Internet network, which is a key component that connects IoT devices, has different players ranging from business subscribers to individual subscribers and from a local network area (LAN) to a worldwide network area (WAN), thereby connecting a wide range of devices and servers [84]. On the one hand, the Internet can provide a wide range of services and applications that can work in synergy with the information collected from sensors to achieve a fully functional IoT system for providing intelligent services. On the other hand, the continuous use of traditional Internet protocol (TCP/IP) to connect billions of objects and devices worldwide is highly vulnerable to numerous security and privacy threats, such as viruses, intrusion and hacking, replay attack and identity theft [15, 48, 85].

*3) Cloud Service Surface*

Cloud computing provides a set of innovative services that are introduced to offer access to stores and processes for obtaining information from anywhere and at any time; accordingly, the requirement for hardware equipment is either limited or eliminated [86]. Cloud computing can be defined as enabling remote access to shared service resources [87, 88]. Cloud computing can serve as a platform that can be used as a base technology to realise the vision of IoT [89]. Cloud computing has significant characteristics that can benefit IoT systems, such as computational and energy efficiencies and storage, service and application over the Internet [89]. The integration of cloud and IoT offers great opportunities for IoT systems. IoT can benefit from the unrestrained resources of the cloud, thereby overcoming the main constraints of IoT, such as computational and energy capabilities [90]. The integration of cloud and IoT offers opportunities for the cloud as well. The cloud can use an IoT device as a bridge to be integrated into real-life applications through a dynamic and distributed means, consequently supplying cloud services to a large consumer base [89, 91]. However, with the integration of cloud and IoT systems, several security concerns arise because such a distributed system is vulnerable to numerous attacks, such as (1) malicious attacks that can manipulate flaws in data security to obtain unauthorised access (e.g. cross-site scripting (XSS), SQL injection flaws, cross-site request forgery (CSRF) and insecure storage [92]); (2) insufficient integrity controls at the data level that can result in security threats by avoiding the authorisation process to directly access the database [92, 93]; and (3) a security threat may exist in all virtualisation software which can be utilised by intruders (e.g. the vulnerability of a virtual server might allow a guest OS to run codes on the host). Consequently, the vulnerability of the virtual server could be exploited to allow the elevation of privileges [92, 93].

Cloud computing has substantial consequences on information privacy and confidentiality. Privacy and confidentiality risks differ significantly according to the terms and conditions between the cloud service provider and the cloud service consumers. However, the integration of IoT devices with cloud computing introduces several privacy concerns, such as exposing highly confidential data (e.g. personal medical data of the holder or home-based sensor data). Privacy is a vital factor that prevents users from adopting IoT devices. Therefore, development should be accompanied with effective privacy protection for a successful IoT system deployment [94-96]. Moreover, multi-tenancy, which is one of the main features of cloud computing, can cause security threats that may lead to private information leakage. Multi-tenancy allows multiple users to store their data using the cloud via the application interface (API). In such a condition, the data of several users can be stored at the same locality, and data in such an environment can be accessed by one of these users. By either hacking through the loop holes in API or inserting the client code into the cloud system, an unauthorised operation attack can be launched against the data [92]. Authorised cloud users might also misuse their permissible access to gain unauthorised privileges and launch attacks, such as internal DoS. Such attacks can be called insider attacks [82], which can introduce a critical trust issue when the cloud is integrated with IoT.

*4) Web and Application Interface*

Most services provided by IoT systems provide users remote access via the Web or mobile applications. For example, in a smart home, the smart things that are connected to home appliances are designed to be controlled by users using their mobile applications or by webpage interfaces in a few cases. Mobile applications have also been developed for smart cars, watches, belts, shoes, glasses, lights, parking and other things that are becoming IoT-based devices controlled by mobile applications. With the rapid development of IoT, virtual and real worlds are being integrated, and soon the difference between the two worlds will become undefinable. IoT devices can interact with one another in real time. This scenario can be ultimately achieved with the help of smartphone applications [97]. Smartphones have become ubiquitous because of the extensive services they provide to users through their applications. Android-based devices are among the popular smart devices. They have captured a massive market because of their open architecture and the popularity of their application programming interface (APIs) among developer groups [98, 99]. However, the open nature of mobile operating systems permits users to download diverse applications involving malicious applications that are uploaded by a third-party to online application stores without thorough security checks [98, 100]. The growing popularity of Android-based devices and other operating system devices has attracted malware developers, followed by a huge increase in Android malware [98]. Malware developers can control smartphones by utilising platform vulnerabilities, extracting private user information or constructing botnets. Furthermore, Android applications may release private information carelessly or maliciously. Consequently, their functioning behaviours, operational models and usage patterns should be recognised to develop practical security methods for mobile devices [98]. Mobile devices are exposed to risks and threats, such as bluesnarfing, bluejacking, eavesdropping, tracking and DoS [15, 75, 101].

*5) New attack surfaces introduced by IoT*

In this section, we discuss new attack surfaces introduced by the IoT environment.

*a)     Threats caused by IoT interdependent environment*

With the rapid growth of IoT objects, the collaboration between objects has become more automated and require less human involvement. IoT objects no longer merely interact with one another like devices within a network. Many IoT devices nowadays are designed to achieve the vision of a smart city, such that many of these devices are controlled by other devices or depend on the operational condition of other devices or the surrounding environment. For example, if a GPS sensor is aware of the traffic situation in a different road from the user's home to work and the user's health condition (asthma) is known, then the GPS should select the route from the user's home to work that is most suitable for his health condition (less traffic and air pollution) based on the health information and traffic and pollution sensors. Similarly, [102] provided another example where a sensor senses that the indoor temperature is raised and a smart plug senses that the air cooler is turned off; then, the windows automatically open. Such interdependent processes are common in applications that utilise IoT devices to achieve a fully automated process. In this environment, the targeted IoT device may be unreachable by an attacker, but the attacker could modify the operation mode of another device or its sensing parameter through the environment that has direct interdependence to launch a threat [102]. Therefore, attacking one surface, such as reducing the temperature or manipulating pollution data, can cause severe effects on other sensors whose operations depend on the information from these sensors. In such an interdependent environment, the attacker can select the weakest nodes in the systems to interrupt the entire systems.

*b)     Interconnected environment*

IoT systems connect billions of devices. This architecture does not only expend the surface of the attack but also the magnitude of the attack. With these densely interconnected devices, an infected thing can become a destructive attack that infects numerous things at a large scale, thus affecting a large part of a city. This situation of nuclear destruction of technology can be described as 'IoT goes nuclear' [103]. Research [103] shows that IoT devices, even with secured industry-standard cryptographic methods, may be exploited by attackers to produce a new-fangled category of security risks that can be circulated from one IoT device to all its physically connected devices through the IoT medium. Consequently, an attacker can launch rapid and destructive attacks that may not be easy to control. To illustrate the impact of this scenario, an experimental case was conducted in [103], where an infection attack was launched by exploiting the popular Philips Hue smart lamps. The malware was diffused by moving directly from one lamp to the adjacent lamp through wireless connectivity provided by the built-in ZigBee and physical proximity. The researchers [103] found that the global AES-CCM key can be used to encrypt and authenticate new firmware without knowing any real updates on smart lamps by using cheap available equipment. This situation shows how vulnerable such devices are, even the devices produced by a well-known company that applies reliable cryptographic methods for security. Such attacks can start at a single point at any location and may end up infecting the entire city, thereby allowing the attackers to control the lights of the city or use IoT lamps in DDoS attacks [103]. Consequently, an infection attack may spread rapidly to large-scale devices and components due to the interconnected nature of IoT systems.

*c)     Social IoT environment*

The Social Internet of Things (SIoT) was introduced recently to integrate social networking into IoT. The basis of such integration is that each thing (object) can obtain preferred services through its social objects called friends in a distributed manner with just local information[104].

Consequently, the threats caused by this IoT environment can be related to privacy concerns that may cause exposing sensitive information about the objects when integrated into a social network of IoT devices [105].

## IV. REVIEW OF MACHINE LEARNING AND DEEP LEARNING APPLICATIONS IN IOT SECURITY

Learning algorithms have been widely adopted in many real-world applications because of their unique nature of solving problems. Such algorithms handle the construction of machines that progress automatically through experience [106]. Recently, learning algorithms have been widely applied in practice. The current advancement of learning algorithms has been driven by the development of new algorithms and the availability of big data, in addition to the emergence of low-computation-cost algorithms [106]. ML and DL have advanced considerably over the past few years, starting from laboratory curiosity and progressing to practical machinery with extensive, significant applications [106]. Even though DL is a ML sub-field, in this paper ML refers to traditional ML methods that require engineered features, while DL methods refer to recent advances in learning methods that utilise several non-linear processing layers for discriminative or generative feature abstraction and transformation for pattern analysis [12]. The purpose of discussing ML and DL in two sections is to provide the readers with in-depth review of both of them.

Generally, learning algorithms aim to improve performance in accomplishing a task with the help of training and learning from experience. For instance, in learning intrusion detection, the task is to classify system behaviour as normal or abnormal. An improvement in performance can be achieved by improving classification accuracy, and the experiences from which the algorithms learn are a collection of normal system behaviour. Learning algorithms are classified into three main categories: supervised, unsupervised and reinforcement learning (RL).

Supervised learning methods form their classification or prediction model on the basis of a learnt mapping [106] and produce by observing the input parameters. In other words, these methods capture the relationships between the input parameters (features) and the required output. Therefore, at the initial stage of supervised learning, learning examples are needed to train the algorithms, which are then used to predict or classify the new input [107]. Recent prodigious advancement in supervised learning engages deep networks. These networks can be viewed as multilayer networks with threshold units [106], each of which calculates the function of its input [108, 109].

Although many practical realisations of DL have originated from supervised learning methods for learning representations, recent works have achieved progress in improving DL systems that learn important representations of the input without the necessity of pre-labelled training data [110]. These learning algorithms are unsupervised learning methods, which are generally intended to analyse unlabelled data. The objective of an unsupervised learning algorithm is to categorise the input data into distinctive groups by examining the similarity between them.

The third common type of ML is RL [111, 112]. RL algorithms are trained by the data from an environment. RL aims to understand an environment and discover the best approaches to a given agent in different environments [113]. The training data in RL are halfway between those of supervised and unsupervised learning. In place of the training samples in which the right output is provided for a specified input, the training data in RL are assumed to indicate whether an action is right or not; if an action is not right, then the problem remains until the right action is discovered [106]. Thus, RL is trial-and-error learning.

In this section, we discuss the most promising ML and DL algorithms in IoT security perspective. Firstly, we discuss the traditional ML algorithms, their advantages, disadvantages and applications in IoT security. Secondly, we discuss DL algorithms, their advantages, disadvantages and applications in IoT security.

### A. Machine learning (ML) methods for IoT security

In this subsection, we discuss the common ML algorithms (i.e. decision trees (DT), support vector machines (SVM), Bayesian algorithms, k-nearest neighbour (KNN), random forest (RF), association rule (AR) algorithms, ensemble learning, k-means clustering and principal component analysis (PCA)) and their advantages, disadvantages and applications in IoT security

### 1) Decision Trees (DTs)

DT-based methods mainly classify by sorting samples according to their feature values. Each vertex (node) in a tree represents a feature, and each edge (branch) denotes a value that the vertex can have in a sample to be classified. The samples are classified starting at the origin vertex and with respect to their feature values. The feature that optimally splits the training samples is deemed the origin vertex of the tree [114]. Several measures are used to identify the optimal feature that best splits the training samples, including information gain [115] and the Gini index [116].

Most DT-based approaches consist of two main processes: building (induction) and classification (inference) [117]. In the building (induction) process, a DT is constructed typically by initially having a tree with unoccupied nodes and branches. Subsequently, the feature that best splits the training samples is considered the origin vertex of the tree. This feature is selected using different measures, such as information gain. The premise is to assign the feature root nodes that maximally reduce the intersection area between classes in a training set, consequently improving the discrimination power of the classifier. The same procedure is applied to each sub-DT until leaves are obtained and their related classes are set. In the classification (inference) process, after the tree is constructed, the new samples with a set of features and unknown class are classified by starting with the root nodes of the constructed tree (i.e. the tree constructed during the training process) and proceeding on the path corresponding to the learnt values of the features at the inner nodes of the tree. This procedure is sustained until a leaf is acquired. Finally, the related labels (i.e. predicted classes) of the new samples are obtained [117].

Researchers in [117] summarised the main points for simplifying DT construction. Firstly, pre-pruning or post-pruning is applied to the tree to reduce the tree size. Secondly, the space of the states searched is adjusted. Thirdly, the search algorithm is enhanced. Next, the data features are reduced by removing or disregarding redundant features through the search process. Finally, the structure of the tree is converted into an alternative data structure, such as a set of rules. The main weaknesses of DT-based methods are summarised as follows [117]. Firstly, they require large storage because of the nature of construction. Secondly, understanding DT-based methods is easy only if few DTs are involved. However, certain applications involve a massive construction of trees and several decision nodes. In these applications, the computational complexity is high, and the underlying model for classifying samples is complex.

A DT is used as a main classifier or collaborative classifier with other ML classifiers in security applications, such as intrusion detection [118, 119]. For example, a previous study proposed the use of a fog-based system call system to secure IoT devices [120]. The research used DT to analyse network traffic to detect suspicious traffic sources and consequently detect DDoS behaviour.

*2) Support Vector Machines (SVMs)*

SVMs are used for classification by creating a splitting hyperplane in the data attributes between two or more classes such that the distance between the hyperplane and the most adjacent sample points of each class is maximised [121]. SVMs are notable for their generalisation capability and specifically suitable for datasets with a large number of feature attributes but a small number of sample points [122, 123]. Theoretically, SVMs were established from statistical learning [121]. They were initially created to categorise linearly divisible classes into a two-dimensional plane comprising linearly separable data points of different classes (e.g. normal or abnormal). SVMs should produce an excellent hyperplane, which delivers maximum margin, by increasing the distance between the hyperplane and the most adjacent sample points of each class. The advantages of SVMs are their scalability and their capabilities to perform real-time intrusion detection and update the training patterns dynamically.

SVMs have been widely used in various security applications, such as intrusion detection [124-126], and are efficient in terms of memory storage because they create a hyperplane to divide the data points with a time complexity equal to $O(N^2)$, where N refers to the number of samples [122, 123]. In relation to the IoT environment, a study [127] developed an Android malware detection system to secure IoT systems and applied a linear SVM to their system. They compared the detection performance of SVM with other ML algorithms, namely, naïve Bayes (NB), RF and DT. The comparison results showed that SVM outperformed the other ML algorithms. Such results confirmed the robust application of SVM for malware detection. Nevertheless, additional studies are essential to investigate the performance of SVMs with enriched datasets and datasets that are created with different environments and attack scenarios. Moreover, comparing the performances of SVM with DL algorithms, such as convolutional neural network (CNN) algorithms, in such a situation may be interesting.

In a previous work, an SVM was used to secure a smart grid, and attack detection in a smart grid was empirically studied [128]. This research showed that the ML algorithms SVM, KNN, perceptron, ensemble learning and sparse logistic regression are effective in detecting known and unknown attacks, performing better than traditional methods used for attack detection in smart grids.

In another research direction, SVM was recently used as a tool to exploit device security. The results in [129, 130] showed that ML methods can break cryptographic devices and that SVM is more effective in breaking cryptographic devices than the traditional method (i.e. template attack).

*3) Bayesian theorem-based algorithms*

Bayes' theorem explains the probability of an incident on the basis of previous information related to the incident [131]. For instance, DoS attack detection is associated with network traffic information. Therefore, compared with assessing network traffic without knowledge of previous network traffic, using Bayes' theorem can evaluate the probability of network traffic being attack (related or not) by using previous traffic information. A common ML algorithm based on Bayes' theorem is the Naive Bayes (NB) classifier.

The NB classifier is a commonly used supervised classifier known for its simplicity. NB calculates posterior probability and uses Bayes' theorem to forecast the probability that a particular feature set of unlabelled examples fits a specific label with the assumption of independence amongst the features. For example, for intrusion detection, NB can be used to classify the traffic as normal or abnormal. The features that can be used for traffic classification, such as connection duration, connection protocol (e.g. TCP and UDP) and connection status flag, are treated by the NB classifier independently despite that these features may depend on one another. In NB classification, all features individually contribute to the probability that the traffic is normal or abnormal; thus, the modifier "naïve" is used. NB have been used for network intrusion detection [132, 133] and anomaly detection [134, 135]. The main advantages of NB classifiers include simplicity, ease of implementation, applicability to binary and multi-class classification, low training sample requirement [136] and robustness to irrelevant features (The features are preserved independently.). However, NB classifiers cannot capture useful clues from the relationships and interactions among features. The interactions among features can be important for accurate classification, particularly in complex tasks in which the interactions among features can significantly help the classifier increase its discrimination power among classes [137].

*4) k-Nearest neighbour (KNN)*

KNN is a nonparametric method. KNN classifiers often use the Euclidean distance as the distance metric [138]. Figure 6 demonstrates KNN classification, in which new input samples are classified. In the figure, the red circles represent malicious behaviours, and the green circles represent the normal behaviours of the system. The newly unknown sample (blue circle) needs to be classified as malicious or normal behaviour. The KNN classifier categorises the new example on the basis of the votes of the selected number of its nearest neighbours; i.e. KNN decides the class of unknown samples by the majority vote of its nearest neighbours. For instance, in Figure 6, if the KNN classification is based on one nearest neighbour (k = 1), then it will categorise the class of the unseen sample as normal behaviour (because the nearest cycle is a green cycle). If the KNN classification is based on two nearest neighbours (k = 2), then the KNN classifier will categorise the class of the unseen sample as normal behaviour because the two nearest circles are green (normal behaviour). If the KNN classification is based on three and four nearest neighbours (k = 3, k = 4), then the KNN classifier will categorise the class of the unknown sample as malicious behaviour because the three and four nearest circles are red circles (malicious behaviour). Testing different values of k during the cross-validation process is an important step to determine the optimal value of k for a given dataset. Although the KNN algorithm is a simple classification algorithm and effective for large training datasets [139], the best k value always varies depending on the datasets. Therefore, determining the optimal value of k may be a challenging and time-consuming process. KNN classifiers have been used for network intrusion detection and anomaly detection [140-145]. Considering the IoT environment, a study [146] proposed a model for the detection of U2R and R2L attacks. The model reduced the dimensionality of the features to enhance efficiency using two layers of feature reduction and then applied a two-tier classification model that uses NB and KNN classifiers; the proposed model showed good detection results for both attacks. Another research developed intrusion detection system-based KNN [147]. The developed system was meant for use in classifying nodes as normal or abnormal in a wireless sensor network (WSN), which is an important unit of IoT systems; the proposed system exhibited efficient and accurate intrusion detection.

### 5) Random forest (RF)

RFs are supervised learning algorithms. In an RF, several DTs are constructed and combined to acquire a precise and robust prediction model for improved overall results [148, 149]. Therefore, an RF consists of numerous trees that are constructed randomly and trained to vote for a class. The most voted class is selected as the final classification output [148]. Even though the RF classifier is constructed mainly using DTs, these classification algorithms substantially differ. Firstly, DTs normally formulate a set of rules when the training set is fed into the network, and this set of rules is subsequently used to classify a new input. RF uses DTs to construct subsets of rules for voting a class; thus, the classification output is the average of the results and RF is robust against over-fitting. Moreover, RF bypasses feature selection and requires only a few input parameters [26]. However, the use of RF may be impractical in specific real-time applications in which the required training dataset is large because RF needs the construction of several DTs. RF algorithms have been used for network intrusion detection and anomaly detection [150, 151]. In a previous study [152], RF, SVM, KNN and ANN were trained to detect DDoS in IoT systems, and RF provided slightly better classification results than did the other classifiers when limited feature sets were used to avoid additional computational overhead and improve the applicability of the system to real-time classification. RF was trained using features obtained from network traffic with the purpose of correctly recognising IoT device categories from the white list. The authors extracted and manually labelled network traffic data from 17 IoT devices. These devices belonged to nine categories of IoT devices and adopted to train a multi-class classifier using RF algorithms. The study concluded that ML algorithms, in general, and specifically RF, hold practical significance in correctly identifying unauthorised IoT devices [153].

### 6) Association Rule (AR) algorithms

AR algorithms [154] have been used to identify an unknown variable by investigating the relationship among various variables in a training dataset. For example, let $X, Y$ and $Z$ be variables in a dataset $T$. An AR algorithm aims to study the relationship among these variables to discover their correlations and consequently construct a model. Subsequently, this model is used to predict the class of new samples. AR algorithms identify frequent sets of variables [26], which are combinations of variables that frequently co-exist in attack examples. For example, in a previous study [155], the associations between TCP/IP variables and attack types were investigated using ARs, and the occurrence of various variables, such as service name, destination port, source port and source IP, were examined to predict the attack type. The AR algorithm reported in [156] exhibited favourable performance in intrusion detection. The researchers used fuzzy association rules in an intrusion detection model, which yielded a high detection rate and a low false positive rate [156]. However, compared with other learning methods, AR methods are not commonly used in IoT environments; thus, further exploration is suggested to check whether an AR method can be optimised or combined with another technique to provide an effective solution to IoT security. The main drawbacks of AR algorithms in practice are as follows. Firstly, the time complexity of AR algorithms is high. Association rules increase rapidly to an unmanageable quantity, particularly when the frequency among variables is decreased. Although several different approaches have been introduced to tackle the issue of efficiency, they are not always effective [157]. Moreover, AR algorithms are based on simple assumptions among variables (direct relationships and occurrence). In certain cases, these assumptions are inapplicable, especially to security applications, in which attackers usually attempt to imitate the behaviour of normal users.

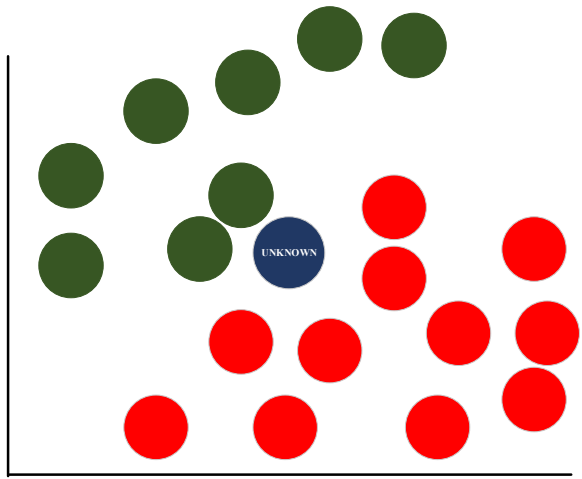

*Figure 6 KNN working principle*

*7) Ensemble learning (EL)*

One of promising directions in ML is EL. EL combines the outputs of numerous basic classification methods to produce a collective output and consequently improve classification performance. EL aims to combine heterogeneous or homogeneous multi-classifiers to obtain a final result [158]. At the initial stage of ML development, every learning method has its advantages and achievements in specific applications or with specific datasets. Experimental comparisons in [159] found that the best learning method differs by application. The underlying learning theory used for a classifier depends on the data. Given that the nature of data apparently changes with the application, the best learning method that suits the given application data may not be the best for other applications. Therefore, researchers have started combining different classifiers to improve accuracy. EL uses several learning methods; thus, it reduces variance and is robust to over-fitting. The combination of different classifiers can provide results beyond the original set of hypotheses; thus, EL can adapt well to a problem [160]. However, the time complexity of an EL-based system is more than that of a single classifier-based system because EL comprises several classifiers [161, 162]. EL has been effectively used for intrusion, anomaly and malware detection [163-166].

A previous study [167] showed that the time complexity of such learning models can be reduced to make them suitable for devices with limited hardware resources, such as IoT devices. the authors proposed a lightweight, application-independent, ensemble learning-based framework for detecting online anomalies in the IoT environment. The proposed framework aims to tackle two issues: 1) accomplishing automated and distributed online learning approaches to identifying anomalies for resource-constrained devices and 2) evaluating the proposed framework with real data. The study reported that the ensemble-based method outperformed each individual classifier [167].

*8) k-Means clustering*

k-Means clustering is based on an unsupervised ML approach. This method aims to discover clusters in the data, and k refers to the number of clusters to be generated by the algorithm. The method is implemented by iteratively allocating each data point to one of the k clusters according to the given features. Each cluster will contain samples with similar features. The k-means algorithm applies iterative refinement to generate an ultimate result. The inputs of the algorithm are the number of clusters (k) and dataset, which contains a set of features for each sample in the dataset. Firstly, the k centroids are estimated, and then each sample is assigned to its closest cluster centroid according to the squared Euclidean distance. Secondly, after all the data samples are assigned to a specific cluster, the cluster centroids are recalculated by computing the mean of all samples assigned to that cluster. The algorithm iterates these steps until no sample that can modify the clusters exists [168, 169]. The main limitations of k-means clustering are as follows. Firstly, the user has to select k in the beginning. Secondly, this algorithm assumes that all spherical clusters have an approximately equal numbers of samples. The k-means algorithms can be applied to anomaly detection by distinguishing normal behaviour from abnormal behaviour by feature similarity calculations[170, 171]. Muniyandi, Rajeswari and Rajaram [172] proposed an anomaly detection method using k-means with DT (i.e. C4.5 DT algorithm). However, the performance of k-means was less effective than those of supervised learning methods, specifically in detecting known attack [173]. Unsupervised algorithms are generally a good choice when generating the labelled data is difficult. However, the application of clustering methods, in general, and k-means in particular, to IoT system security is still at its infancy and should be explored further.

Unsupervised ML methods have many applications in securing IoT systems. For instance, k-means clustering was used for securing WSNs by detecting intrusions [174]. In a study on Sybil detection in industrial WSNs [175], a kernel-oriented scheme was proposed to differentiate Sybil attackers from normal sensors by clustering the channel vectors. A clustering algorithm showed the potential to preserve private data anonymisation in an IoT system [176]. The use of clustering to develop data anonymisation algorithms can significantly advance data exchange security [176].

*9) Principal component analysis (PCA)*

PCA is a feature-reduction technique that can be applied to transform a large set of variables into a reduced set that preserves most of the information represented in the large set. This technique converts a number of probably correlated features into a reduced number of uncorrelated features, which are called principal components [177]. Therefore, the main working principle of PCA can be utilised for feature selection to realise real-time intrusion detection for IoT systems; a previous work proposed a model that uses PCA for feature reduction and adopts softmax regression and KNN algorithm as classifiers. The author reported that the combination of PCA with these classifiers provided a time- and computing-efficient system that can be utilised in real time in IoT environments [178].

Table 1 shows Potential ML methods for securing IoT systems and their advantages, disadvantages and applications in IoT security.

*Table 1* Potential ML methods for securing IoT system

| Method | Working principle | Advantages | Disadvantages | Potential Application in IoT Security |
|---|---|---|---|---|
| **DT** | DT-based method uses a DT to establish a model (i.e. a prediction model) to learn from training samples by representing them as branches and leaves. The pre-trained model is then used to predict the class of the new sample. | DT is a simple, easy-to-use and transparent method. | DT requires large storage because of its construction nature. Understanding DT-based methods is easy only if few DTs are involved. | Detection of intrusion [118, 119] and suspicious traffic sources [120] |
| **SVM** | SVMs form a splitting hyperplane in the feature dimension of two or more classes such that the distance between the hyperplane and the most adjacent sample points of each class is maximised [121]. | SVMs are known for their generalisation capability and suitability for data consisting of a large number of feature attributes but a small number of sample points [122, 123]. | The optimal selection of a kernel is difficult. Understanding and interpreting SVM-based models are difficult. | Detection of intrusion [124-126], malware [127] and attacks in smart grids [128] |
| **NB** | NB calculates the posterior probability. It uses Bayes' theorem to forecast the probability that a particular feature set of unlabelled samples fits a specific label with the assumption of independence amongst features. | NB is known for its simplicity, ease of implementation, low training sample requirement [136] and robustness to irrelevant features (The features are preserved independently.). | NB handles features independently and thus cannot capture useful clues from the relationships and interactions among features. (It may work effectively in applications whose samples have dependent and related features.) | Detection of network intrusion [132, 133]. |
| **KNN** | KNN classifies the new sample on the basis of the votes of the selected number of its nearest neighbours; i.e. KNN decides the class of unknown samples by the majority vote of its nearest neighbours. | KNN is a popular and effective ML method for intrusion detection. | The optimal $k$ value usually varies from one dataset to another; therefore, determining the optimal value of $k$ may be a challenging and time-consuming process. | Detection of intrusions [146] and anomalies [140-145]. |
| **RF** | In an RF, several DTs are constructed and combined to acquire a precise and established prediction model for improved overall results. | RF is robust to over-fitting. RF bypasses feature selection and requires only a few input parameters. | RF is based on constructing several DTs; thus, it may be impractical in specific real-time applications in which the required training dataset is large. | Detection of intrusion [150], anomalies [151], DDoS attacks [152] and unauthorised IoT devices [153] |
| **AR algorithm** | AR algorithms aim to study the relationship among the variables in a given training dataset *T* to discover correlations and consequently construct a model. This model is then used to predict the class of new samples. | AR algorithms are simple and easy to use. | The time complexity of the algorithms is high. AR algorithms use simple assumptions among variables (direct relationships and occurrence). In certain cases, these assumptions are inapplicable, especially to security applications. | Detection of intrusion [156] |
| **EL** | EL combines the outputs of numerous basic classification methods to produce a collective output and consequently improve classification performance. | EL reduces variance and is robust to over-fitting. EL provides results beyond the original set of hypotheses; therefore, EL can adapt better than can a single classifier-based method to a problem. | The time complexity of an EL system is higher than that of a single classifier-based system. | Detection of intrusion, anomalies and malware [163-167]. |
| ***k*-Means clustering** | *k*-Means clustering is an unsupervised learning approach that identifies clusters in the data according to feature similarities. *k* refers to the number of clusters to be generated by the algorithm. | Unsupervised algorithms are generally a good choice when generating the labelled data is difficult. *k*-Means clustering can be used for private data anonymisation in an IoT system because it does not require labelled data. | k-Means clustering is less effective than supervised learning methods, specifically in detecting known attacks [173]. | Sybil detection in industrial WSNs [175] and private data anonymisation in an IoT system [176] |
| **PCA** | PCA is a process that converts a number of probably correlated features into a reduced number of uncorrelated features, which are called principal components [177]. | PCA can achieve dimensionality reduction and consequently reduce the complexity of the model. | PCA is a feature-reduction method that should be used with other ML methods to establish an effective security approach. | PCA can be used for real-time detection systems in IoT environments [178] by reducing the model features. |

## B. Deep learning (DL) methods for IoT Security

Recently, the applications of DL to IoT systems have become an imperative research topic [179]. The most vital advantage of DL over traditional ML is its superior performance in large datasets. Several IoT systems produce a large amount of data; thus, DL methods are suitable for such systems. Moreover, DL can automatically extract complex representations from data [179]. DL methods can enable the deep linking of the IoT environment [180]. Deep linking is a unified protocol that permits IoT-based devices and their applications to interact with one another automatically without human intervention. For example, the IoT devices in a smart home can automatically interact to form a fully smart home [179].

DL methods provide a computational architecture that combines several processing levels (layers) to learn data representations with several levels of abstraction. Compared with traditional ML methods, DL methods have considerably enhanced state-of-the-art applications [12]. DL is a ML sub-field that utilises several non-linear processing layers for discriminative or generative feature abstraction and transformation for pattern analysis. DL methods are also known as hierarchical learning methods because they can capture hierarchical representations in deep architecture. The working principle of DL is inspired by the working mechanisms of the human brain and neurons for processing signals. Deep networks are constructed for supervised learning (discriminative), unsupervised learning (generative learning) and the combination of these learning types, which is called hybrid DL. CNNs and recurrent neural networks (RNNs) are examples of discriminative DL methods. Deep autoencoders (AEs), deep belief networks (DBN), restricted Boltzmann machines (RBMs), generative adversarial networks (GANs) and ensemble of DL networks (EDLNs) are examples of hybrid DL methods.

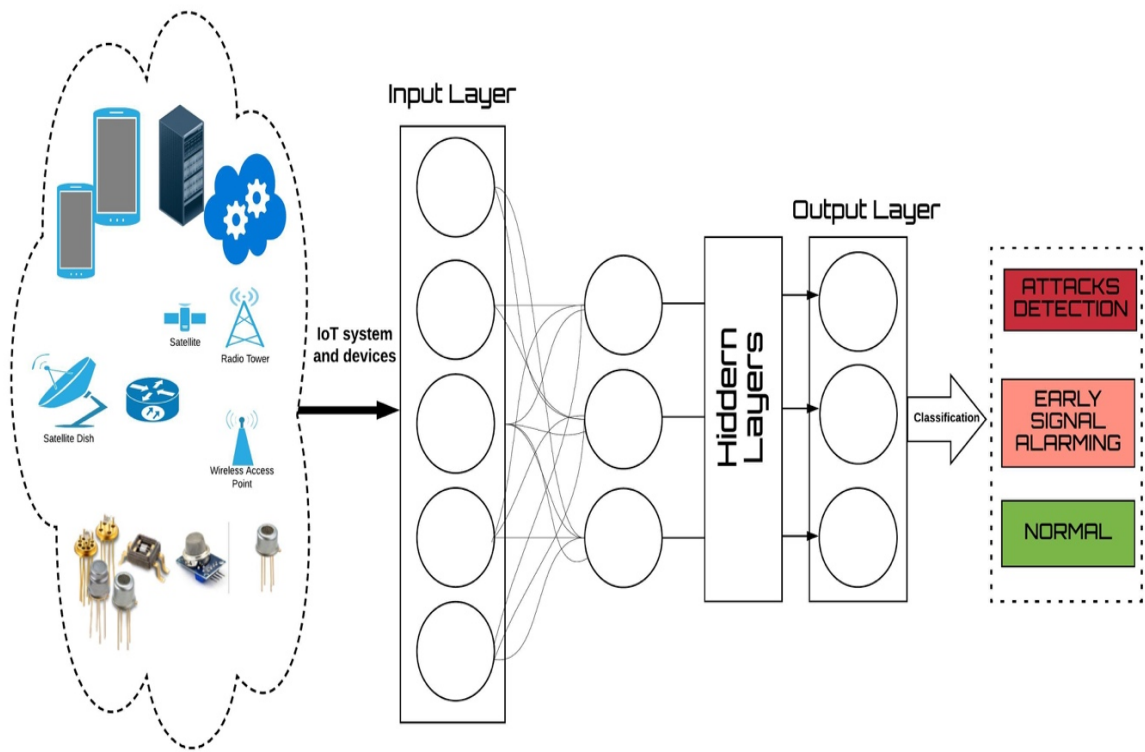

*Figure 7 Illustration of NNs Working Principle for IoT Security*

### 1) Convolutional neural networks (CNNs)

CNNs were introduced to reduce the data parameters used in a traditional artificial neural network (ANN). The data parameters are reduced by utilising three concepts, namely, sparse interaction, parameter sharing and equivariant representation [181]. Reducing the connections between layers increases the scalability and improves the training time complexity of a CNN.

A CNN consists of two alternating types of layers: convolutional layers and pooling layers. The convolutional layers convolute data parameters with the help of multiple filters (kernels) of equal size [11]. The pooling layers perform down-sampling to decrease the sizes of the subsequent layers through max pooling or average pooling. Max pooling divides the input into non-overlapping clusters and selects the maximum value for each cluster in the previous layer [182, 183], whereas average pooling averages the values of each cluster in the previous layer. Another important layer of a CNN is the activation unit, which performs a non-linear activation function on each element in the feature space. The non-linear activation function is selected as the rectified linear unit (ReLU) activation function, which involves nodes with the activation function $f(x) = max(0, x)$ [184]. The working principle of CNN applied to IoT Security is shown in Figure 8.

The main advantage of a CNN is that it is extensively applied to the training approaches in DL. It also allows for the automatic learning of features from raw data with high performance. However, a CNN has high computational cost; thus, implementing it on resource-constrained devices to support on-board security systems is challenging. Nevertheless, distributed architecture can solve this issue. In this architecture, a light deep neural network (DNN) is implemented and trained with only a subset of important output classes on-board, but the complete training of the algorithm is achieved at cloud level for deep classification [185].

The development of CNNs is mainly directed towards image recognition advancement. Accordingly, CNNs have become widely used, leading to developing successful and effective models for image classification and recognition with the use of large public image sources, such as ImageNet [186, 187]. Furthermore, CNNs demonstrate robustness in numerous other applications. For IoT security, a study [188] proposed a CNN-based malware detection method for Android. With the application of the CNN, the significant features related to malware detection are learnt automatically from the raw data, thereby eliminating the need for manual feature engineering. The key point in using a CNN is that the network is trained to learn suitable features and execute classification conjointly, thus eliminating the extraction process required in traditional ML and consequently providing an end-to-end model [188]. However, the robust learning performance of CNNs can be used by attackers as a weapon. A previous study [189] showed that a CNN algorithm can break cryptographic implementations successfully.

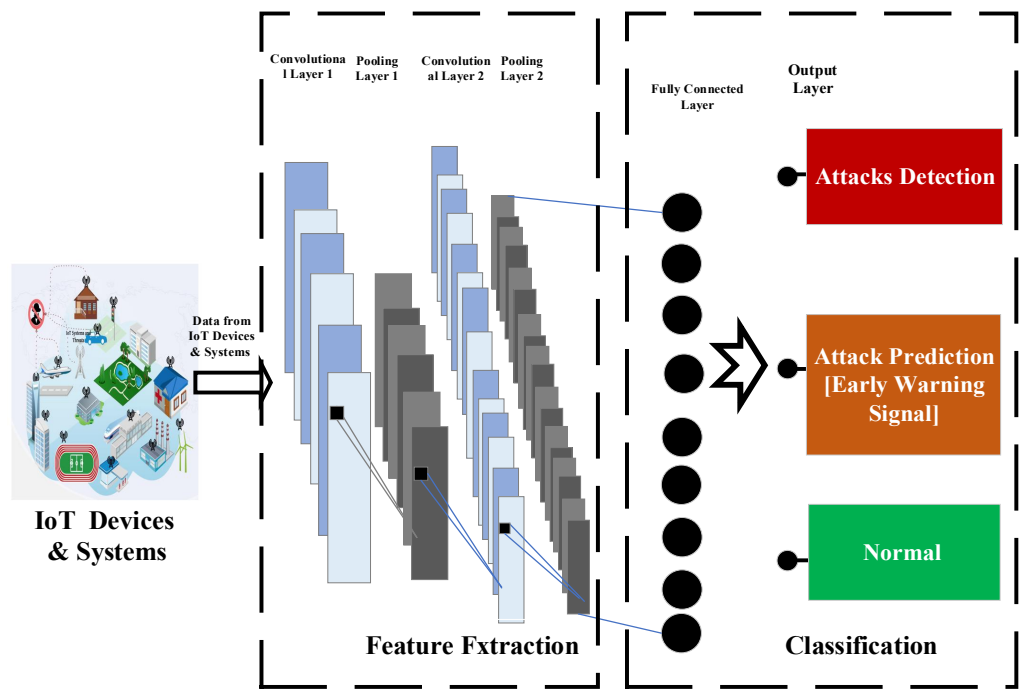

*Figure 8 Illustration of CNN Working Principle for IoT Security*

### 2) Recurrent neural networks (RNNs)

An RNN is a vital category of DL algorithms. RNNs were proposed to handle sequential data. In several applications, forecasting the current output is based on the analysis of the associations from several previous samples. Thus, the output of the neural network depends on the present and past inputs. In such an arrangement, a feed-forward NN is inappropriate because the association between the input and output layers are preserved with no dependency [190]. Therefore, when the backpropagation algorithm was introduced, its most remarkable application was the training of RNNs [12, 191]. For applications that consist of sequential inputs (e.g. speech, text and sensor data), RNNs are recommended [12, 191].

An RNN integrates a temporal layer to capture sequential data and then learns multifaceted variations through the hidden units of the recurrent cell [192]. The hidden units are modified according to the data presented to the network, and these data are continually updated to reveal the present condition of the network. The RNN processes the present hidden state by estimating the subsequent hidden state as an activation of the formerly hidden state. RNNs are used because of their capability of managing sequential data effectively. This capability is advantageous for various tasks, such as threat detection, in which the patterns of the threat are time dependent. Therefore, using recurrent connections can improve neural networks and reveal important behaviour patterns. The main drawback of RNNs, however, is the issue of vanishing or exploding gradients [193].

RNNs and their variants have achieved excellent performance in many applications with sequential data, such as machine translation and speech recognition [194-196]. Moreover, RNNs can be used for IoT security. IoT devices generate large amounts of sequential data from several sources, such as network traffic flows, which are among the key features for detecting several potential network attacks. For example, a previous study [197] discussed the feasibility of an RNN in examining network traffic behaviour to detect potential attacks (malicious behaviour) and confirmed the usefulness of the RNN in classifying network traffic for accurate malicious behaviour detection. Thus, RNNs provide a practical solution in real-world scenarios. Exploring RNNs and their variants are of significance in improving IoT system security, specifically for time series-based threats.

*3) Deep autoencoders (AEs)*

A deep AE is an unsupervised learning neural network trained to reproduce its input to its output. An AE has a hidden layer h, which defines a code used to represent the input [181]. An AE neural network is divided into two parts: the encoder function $h = f(x)$ and the decoder function, which attempts to reproduce the input $r = g(h)$. The encoder obtains the input and converts it into an abstraction, which is generally termed as a code. Subsequently, the decoder acquires the constructed code, which was initially produced to represent the input, to rebuild the original input. The training process in AEs should be accomplished with minimum reconstruction error [198]. However, AEs cannot learn to replicate the input perfectly. AEs are also restricted because they can produce an approximate copy only, merely copying the inputs that are similar to the training data. The model is required to prioritise which characteristics of the inputs should be copied; thus, it frequently learns useful characteristics of the data [181]. AEs are potentially important for feature extraction. AEs can be successfully used for representation learning to learn features (in place of the manually engineered features used in traditional ML) and reduce dimensionality with no prior data knowledge. AEs, nevertheless, consume high computational time. Although AEs can effectively learn to capture the characteristics of the training data, they may only complicate the learning process rather than represent the characteristics of the dataset if the training dataset is not representative of the testing dataset.

AEs were used to detect network-based malware in [199]; the AEs were trained to learn the latent representation of a diverse feature set; particularly, AEs were trained on the feature vector extracted from the cybersystems. The AEs exhibited better detection performance than did the traditional ML algorithms SVM and KNN [199]. In another study [200], an AE was combined with a DBN to construct a malware detection method and used for data dimensionality reduction by non-linear mapping to extract only the significant features; subsequently, the DBN learning algorithm was trained to detect malicious code.

*4) Restricted Boltzmann machines (RBMs)*

RBMs are deep generative models developed for unsupervised learning [201]. An RBM is a completely undirected model with no link between any two nodes in the same layer. RBMs consist of two types of layers: visible and hidden layers. The visible layer holds the known input, whereas the hidden layer consists of multiple layers that include the latent variables. RBMs hierarchically understand features from data, and the features captured in the initial layer are used as latent variables in the following layer.

The research in [202] developed a network anomaly detection model that can overcome the inherent challenges in developing such a model. These challenges include the generation of labelled data required for the effective training of the model because a network traffic dataset is multi-part and irregular. The second challenge is the constant evolution of anomaly behaviour with time. Therefore, the model should be dynamically adapted to detect any new form of attacks and generalised to detect the anomaly in different network environments. To solve these challenges, the researchers in [202] proposed a learning model that is based on a discriminative RBM, which they selected due to its capability to combine generative models with suitable classification accuracy to detect network anomaly in a semi-supervised fashion even with incomplete training data. However, their experimental results showed that the classification performance of the discriminative RBM was affected when the classifier was tested on a network dataset that differed from the network dataset on which the classifier was trained. This finding should be further investigated, and how a classifier can be generalised to detect an anomaly in different network environments should be further studied.

The feature representation capability of a single RBM is limited. However, RBM can be substantially applied by stacking two or more RBMs to form a DBN. This process is discussed in the following section.

*5) Deep belief networks (DBNs)*

DBNs are generative methods [203]. A DBN consists of stacked RBMs that execute greedy layer-wise training to accomplish robust performance in an unsupervised environment. In a DBN, training is accomplished layer by layer, each of which is executed as an RBM trained on top of the formerly trained layer (DBNs are a set of RBMs layers used for the pre-training phase and subsequently become a feed-forward network for weight fine-tuning with contrastive convergence.) [192]. In the pre-training phase, the initial features are trained through a greedy layer-wise unsupervised approach, whereas a softmax layer is applied in the fine-tuning phase to the top layer to fine-tune the features with respect to the labelled samples [195].

DBNs have been successfully implemented in malicious attack detection. A previous study [204] proposed an approach to secure mobile edge computing by applying a DL-based approach to malicious attack detection. The study used a DBN for automatic detection, and the proposed DBN-based model showed vital improvement in malware detection accuracy compared with ML-based algorithms [204]. This result demonstrated the superiority of DL, in general, and DBNs in particular, to traditional manual feature engineering methods in malware detection. In another study [200], an AE was combined with a DBN to establish a malware detection method, and an AE DL algorithm was used for the reduction of data dimensionality by non-linear mapping to extract only the significant features; subsequently, the DBN learning algorithm was trained to detect malicious code.

DBNs are unsupervised learning methods trained with unlabelled data iteratively for significant feature representation. However, even though DBNs use contrastive convergence to reduce computational time, these networks are still inapplicable to on-board devices with limited resources.

*6) Generative adversarial networks (GANs)*

Introduced by [205], GANs have recently emerged as promising DL frameworks. A GAN framework simultaneously trains two models, namely, generative and discriminative models, via an adversarial process as shown in Figure 9. The generative model learns the data distribution and generates data samples, and the discriminative model predicts the possibility that a sample originates from the training dataset rather than the generative model (i.e. evaluates the sample for authenticity). The objective of training the generative model is to increase the probability that the discriminative model misclassifies the sample [205]. In each stage, the generative model, which is the generator, is prepared to deceive the discriminator by generating a sample dataset from random noise. By contrast, the discriminator is fed with several real data samples from the training set, accompanied by the samples from the generator. The discriminator aims to classify real (from the training dataset) and unreal (from the generative model) samples. The performances of the discriminative and generative models are measured by the correctly and incorrectly classified samples, respectively. Subsequently, both models are updated for the next iteration. The output discriminative model assists the generative model to enhance the samples generated for the subsequent iteration [198].

GANs have been recently implemented in IoT security. For example, the study in [206] proposed an architecture for securing the cyberspace of IoT systems, and the proposed architecture involves training DL algorithms to classify the system behaviour as normal or abnormal. GAN algorithms were integrated into the proposed architecture for preliminary study, whose evaluation results showed the effectiveness of the GAN-based architecture in detecting abnormal system behaviour [206].

GANs may have a potential application in IoT security because they may learn different attack scenarios to generate samples similar to a zero-day attack and provide algorithms with a set of samples beyond the existing attacks. GANs are suitable for training classifiers through a semi-supervised approach. GANs can generate samples more rapidly than can fully visible DBNs because the former is not required to generate different entries in the samples sequentially. In GANs, generating a sample needs only one pass through the model, unlike in RBMs, which require an unidentified number of iterations of a Markov chain [205, 207]. However, GAN training is unstable and difficult. Learning to generate discrete data, such as text, by using a GAN is a challenging task [205, 207].

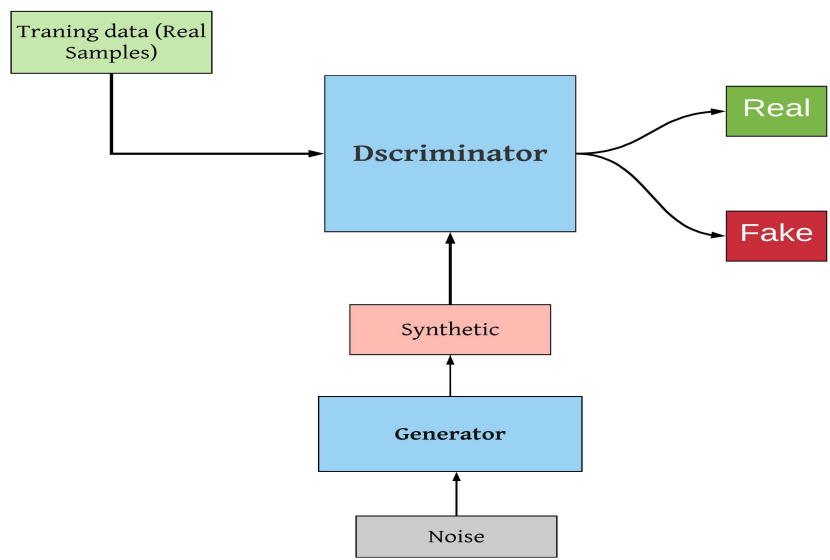

*Figure 9 Illustration of GAN Working Principle*

*7) Ensemble of DL networks (EDLNs)*

Several DL algorithms can work collaboratively to perform better than independently implemented algorithms. EDLNs can be accomplished by merging generative, discriminative or hybrid models. EDLNs are often used to handle complex problems with uncertainties and high-dimensional features. An EDLN comprises stacked individual classifiers, either homogenous (classifiers from the same family) or heterogeneous (classifiers from different families), and is used to enhance diversity, accuracy, performance and generalisation [208]. Although EDLNs have achieved remarkable success in many applications, such as human activity recognition, EDLNs application in IoT security needs further investigation, particularly the possibility of implementing light homogenous or heterogeneous classifiers in a distributed environment to improve the accuracy and performance of an IoT security system and solve challenges related to computational complexity. Table 2 shows Potential DL methods for securing IoT systems and their advantages, disadvantages and applications in IoT security.

*Table 2* Potential DL methods for securing IoT systems

| Methods | Working principle | Advantages | Disadvantages | Potential Application in IoT Security |
|---|---|---|---|---|
| CNNs | CNNs mainly aim to reduce data parameters used by applying sparse interactions, parameter sharing and equivariant representations [181], thereby reducing the connections between layers to quantities less than those in ANNs. | CNNs are robust supervised DL methods with highly competitive performance. With the new features of CNNs, their scalability is increased and their training time complexity is improved compared with those of ANNs. CNNs have potential application in IoT security as they can automatically learn features from security raw data. | CNNs have high computational cost; thus, implementing them on resource-constrained devices to support on-board security systems is challenging. | Malware detection [188]; CNNs can automatically learn features of raw security data; therefore, they can construct an end-to-end security model for IoT systems [188]. |
| RNNs | RNNs integrate a temporal layer to take sequential data and then learn multi-faceted variations with the hidden unit of the recurrent cell [192]. | RNNs and their variants have achieved excellent performance in many applications with sequential data. In certain cases, IoT security data consist of sequential data; thus, RNNs have potential application in IoT security. | The main drawback of RNNs is the issue of vanishing or exploding gradients [193]. | RNNs can classify network traffic with high accuracy in detecting malicious behaviour [197]. RNNs and their variants show considerable potential in improving IoT system security, specifically for time series-based threats. |
| AEs | An AE has a hidden layer $h$, which has a code to represent the input. An AE neural network is divided into two parts: the encoder function $h = f(x)$ and the decoder function, which attempts to reproduce the input $r = g(h)$. The encoder obtains the input and converts it into an abstraction, which is generally termed as a code. Subsequently, the decoder acquires the constructed code that was initially produced to represent the input to rebuild the original input. | AEs are potentially important for feature extraction. AEs can be effectively used for representation learning to learn features in place of the manually engineered features used in traditional ML and reduce dimensionality with no prior data knowledge. | AEs consume considerable computational time. Although AEs can effectively learn to capture the characteristics of the training data, if the training dataset is not representative of the testing dataset, then the AEs may only complicate the learning process rather than represent the characteristics of the dataset. | AEs can be used for detecting malware [199]. AE has been combined with a DBN to establish a malware detection method [200]. |
| RBMs | RBMs are deep generative models developed for unsupervised learning [201]. They are completely undirected models with no link between any two nodes in the same layer. | Using a feedback mechanism on RBMs allows for the extraction of numerous vital features through an unsupervised approach. | RBMs have high computational cost; thus, implementing them on resource-constrained IoT devices to support on-board security systems is challenging. | RBMs can be used for network anomaly detection [202]. |
| DBNs | DBNs consist of stacked RBMs that execute greedy layer-wise training to accomplish robust performance in an unsupervised environment. | DBNs are unsupervised learning methods trained with unlabelled data iteratively for significant feature representation. | DBNs present high computational cost due to the extensive initialisation process caused by the large number of parameters. | DBNs can be used for malicious attack detection [204]. |
| GANs | The GAN framework simultaneously trains two models (i.e. generative and discriminative models) via an adversarial process. The generative model learns the data distribution and generates data samples, and the discriminative model predicts the possibility that a sample originates from the training dataset rather than the generative model (i.e. evaluates the instance for authenticity). | In GANs, generating a sample needs only one pass through the model, unlike in DBNs and RBMs in which an unidentified number of iterations of a Markov chain is required [205, 207]. | GAN training is unstable and difficult. Learning to generate discrete data by using GAN is a difficult task [205, 207]. | GANs can be used to build an architecture for securing the cyberspace of IoT systems [206]. |
| EDLNs | EDLNs can be accomplished by merging generative, discriminative or hybrid models. | Combining DL classifiers can help achieve model diversity, improve model performance and expand model generalisation. | The time complexity of the system can be significantly increased. | The use of GANs in securing IoT systems needs further investigation, particularly the possibility of implementing light homogenous or heterogeneous classifiers in a distributed environment to improve the accuracy and performance of a system. |

## C. Reinforcement learning (RL) methods for IoT security

Learning from the surrounding environment is one of the first learning methods humans experience. Humans naturally start learning by interacting with their environment. RL is inspired by the psychological and neuroscientific perspectives on animal behaviour and of the mechanism by which agents can enhance their control of the environment [111, 112]. RL involves making an agent learn how to map situations to actions appropriately to achieve the highest rewards [112]. The agent does not have previous knowledge of which actions to implement but has to learn which actions produce the most rewards by attempting them through trial and error. The features 'trial' and 'error' are the main and unique features of RL. Thus, the agent continues to learn from its experience to increase its rewards. One of the recent successful RL methods is the deep Q network [111]. Extensions of deep Q networks have been suggested, including double Q-learning [209], continuous control with deep RL [210] and prioritised experience replay [211].

RL has been implemented to solve several IoT issues. Studies by [212, 213] proposed an anti-jamming scheme that is based on reinforcement learning for wideband autonomous cognitive radios (WACRs). In [212], information about sweeping jammer signal and unintentional interference was used to distinguish it from other WACRs; RL was used with this information to learn a sub-band selection policy accurately to evade the jammer signal and interference from other WACRs. Similarly, in [213], an RL method based on Q-learning was trained to effectively avoid jamming attacks sweeping over a wide spectrum of hundreds of MHz in real time. In the same direction, [214] used RL to develop an anti-jamming scheme for cognitive radios and integrate the scheme with deep CNN to improve the efficiency of RL in a large number of frequency channels. A similar scheme against aggressive jamming was proposed using deep RL in [215], in which jamming was considered activated in an aggressive environment, which is normally expected in tactical mobile networking; the results showed that RL is a promising method of developing schemes against aggressive jamming.

## V. IOT SECURITY LAYERS BASED ON ML AND DL METHODS

In this section, we classify the previous studies on ML and DL methods for IoT security according to the layers these methods intend to protect. Even though ML and DL may be applied to protect more than one layer or the end-to-end system (which is the advantage of ML and DL methods over other methods and holds potential future uses), the following classification is proposed to highlight the conceptualisation of the ML and DL methods for IoT security. At the end of this section the technology tools that can essentially enable ML/DL deployment for IoT security are listed.

### A. Perception layer

One of the promising applications of DL methods is physical-layer authentication. Traditional physical-layer authentication techniques apply assumption checks and relate the randomness and exclusiveness of the radio channel between "Alice" and "Bob", to detect spoofing attacker "Eve" in a wireless network. Nonetheless, such an approach is not always practical, specifically in dynamic networks [216]. Wang, Jiang, Lv and Xiao [216] used a learning model to construct a physical-layer authentication model that uses past data generated from a spoofing model as learning vectors to train an extreme learning machine. The proposed model exhibited improved spoofing detection performance and consequently achieved considerably enhanced authentication accuracy compared with that of state-of-the-art methods.

Shi, Liu, Liu and Chen [217] proved that the present Wi-Fi signals generated by IoT objects can be adopted to detect distinctive human behavioural and physiological features and can be utilised to authenticate individuals on the basis of an understanding of their daily activities. The authors proposed a scheme which adopts a single pair of Wi-Fi signals generated by IoT devices to mine Wi-Fi channel state information and thus obtain the amplitude and the relative phase for precise user authentication without the need for user participation. Using these features, the authors developed a DL model (i.e. Deep Neural Network (DNN)) to identify the daily human activity distinctiveness of each individual and subsequently generate a fingerprint for each user, called Wi-Fi fingerprint, to capture the distinct characteristics of different users; the proposed DL-based authentication method exhibited high accuracy [217]. This study validates the potential application of DL algorithms in constructing authentication systems.

In another study [215], a scheme against aggressive jamming was developed using RL, and jamming was considered activated in an aggressive environment, which is normally expected in tactical mobile networking. RL was found effective in developing a method against aggressive jamming [215].

The research in [218] also considered the issue of jamming in an IoT network and introduced a centralised approach to addressing possible jamming attacks in an IoT environment, which consists of resource-constrained devices. The idea of the proposed model is to use the IoT access point to protect against the jamming attacker by distributing its power over the sub-carriers in an intelligent manner and using an evolutionary-based algorithm. The proposed method can converge in a practical iteration number; thus, it can provide a better solution than a random power allocation strategy.

Along the same direction, two previous studies [212, 213] proposed RL-based anti-jamming schemes for WACRs. In [212], the authors used information about sweeping jammer signal and unintentional interference to distinguish it from those of other WACRs. This information and RL were combined to learn a sub-band selection policy accurately to evade the jammer signal and interference from other WACRs. Similarly, in [213], an RL method based on Q-learning was trained to effectively avoid jamming attacks sweeping over a wide spectrum of hundreds of MHz in real-time. In the same direction, [214] used RL to develop an anti-jamming scheme for cognitive radios and integrate it with deep CNN to improve the efficiency of RL in a large number of frequency channels.

Incorporating cognitive radio (CR) capability into IoT devices has paved the way for an innovative research on IoT systems [219]. Currently, many researchers are conducting studies on communication and computing in IoT systems. According to two previous studies [220, 221], IoT systems cannot be sustained without comprehensive cognitive capability because of growing issues. CRs are radio devices that can learn and change in accordance with their dynamic environment [222]. The main step towards accomplishing such cognitive operation is enabling CRs to sense and understand their working environment. Ideally, CRs should be able to work over a wide frequency range. However, sensing all required frequencies in real time is a challenging task, specifically with the existence of jamming attacks. CRs can become increasingly useful and reliable communication systems if they can eliminate the incidence of accidental interference or deliberate jamming attacks [213].

B. *Network layer*

The network layer forms the largest surface of the IoT system. This layer is responsible for transmitting and routing data. It provides a ubiquitous access environment to the perception layer, i.e. data communication and storage functionalities [223]. Therefore, securing the IoT network layer should be of high technical priority. Along the same line of thought, Yavuz [224] proposed a DL-based model to detect the routing protocol for IoT systems and created a dataset for training and testing the DL model by using the Cooja IoT simulator with simulations up to 1000 nodes within 16 networks to detect three types of attacks, namely decreased rank attack, hello flood attack and version number attack. The DNN achieved high performance in detecting the three attacks. However, the authors did not mention the statistics as to how many normal and anomalous samples were in the created dataset. Precision, recall and f-measure were used as evaluation metrics; however, in model evaluation, they may not reflect the actual performance of the model and tend to be biased if the created dataset is imbalanced.

Nobakht, Sivaraman and Boreli [225] proposed an intrusion detection framework that is implemented at the network level and constructed ML algorithms to protect smart devices installed in home environments. They used precision and recall metrics to measure the performance of the classifiers. However, the dataset used was unbalanced, with the number of illegal access samples forming the majority of the samples in the dataset; thus, both evaluation metrics may not precisely reflect the model performance. In case of imbalanced data, other performance metrics such as the area under the receiver operating characteristic curve (AUC) can be better choice to evaluate the performance than accuracy, recall and precision metrics [226, 227].

A previous study [197] discussed the viability of an RNN (i.e. large short-term memory [LSTM] network) in the analysis of network traffic behaviour to detect potential attacks (malicious behaviour) and confirmed the effectiveness of the RNN in precisely classifying network traffic to detect malicious behaviour; thus, the LSTM network can be adopted as a practical solution in real-world scenarios.

Cañedo and Skjellum [228] used ML to detect anomalies, specifically training ANN algorithms to detect whether the data sent from an edge to the smart object in an IoT system are valid or invalid. They generated the data from the edge to the device nodes and then inserted invalid and valid data to train the model; the experimental results showed that the ANN can effectively detect invalid data. However, diverse and enriched datasets that contain various data tampering attacks should be used to train and test the ANN to reconfirm whether it can maintain high accuracy in practical settings or other advanced learning algorithms are required. An investigation in this research direction is recommended to generate enriched datasets.

In another study [229], an intrusion detection system (IDS) based on a hybrid detection method (i.e. unsupervised ML

25method with specification-based method) was used for an IoT system. For this purpose, the author proposed a local intrusion detection method at the local node by using a specification-based intrusion detection approach; the method examined the behaviour of the host nodes and sent analysis results to the global node, which used an ML-based intrusion detection method (i.e. unsupervised optimum-path forest algorithm [230] for clustering the data from local node on the basis of the MapReduce design [231]).

A generative model (i.e. unsupervised model) using AEs was proposed in [199] to detect malware network-based anomaly in cybersystems. The AEs were trained to learn the latent representation of a diverse feature set, and they received a feature vector extracted from the cybersystems; compared with SVM and KNN, the AEs exhibited improved detection performance [199].

Wi-Fi technology is an IoT-enabling technology, especially for smart homes [232]. Wi-Fi technology is of practical importance to the expansion of IoT [233]. A previous study [234] aimed to detect impersonation attacks in a Wi-Fi environment by developing a method called weighted feature selection for extracting and selecting deep features, which were combined with the features generated by a stacked AE (SAE) algorithm. The combined features were then fed into a neural network to train it for classifying the input data into two classes (i.e. impersonation or normal) [234]. This combination of unsupervised DL algorithm (i.e. SAE) and supervised DL algorithm (i.e. ANN) showed high detection accuracy, confirming the potential applications of deep algorithms in securing Wi-Fi networks from impersonation attacks. A similar study [235] used a combination of two unsupervised algorithms (SAE) for mining features and k-means clustering for categorising the input into two classes: benign and malicious.

The research in [202] developed a network anomaly detection model that can overcome the inherent challenges in developing such a model. These challenges include the generation of labelled data required for the effective training of the model because a network traffic dataset is multi-part and irregular. The second challenge is the constant evolution of anomaly behaviour with time. Therefore, the model should be dynamically adapted to detect any new form of attacks and generalised to detect the anomaly in different network environments. To solve these challenges, the researchers in [202] proposed a learning model that is based on a discriminative RBM, which they selected due to its capability to combine generative models with suitable classification accuracy to detect network anomaly in a semi-supervised fashion even with incomplete training data. However, their experimental results showed that the classification performance of the discriminative RBM was affected when the classifier was tested on a network dataset that differed from the network dataset on which the classifier was trained. This finding should be further investigated, and how a classifier can be generalised to detect an anomaly in different network environments should be further studied.

Saied, Overill and Radzik [236] used an ANN to detect known and unknown DDoS attacks in a real-time environment. The proposed defence technique aimed to thwart fake packets and permit real packets to pass through. They assessed the ANN's performance in unknown DDoS detection when it is trained with old and updated datasets and reported that the further they trained the algorithm with the latest features of known DDoS attacks, the more they improved the detection probabilities for known and unknown DDoS attacks. The ANN algorithm learns from training samples and then detects zero-day attack features, which are comparable to the features on which it was trained [236].

Chen, Zhang and Maharjan [204] developed a DL-based model for malicious attack detection to secure mobile edge computing. The approach used a DBN for automatic detection, and the model exhibited improved accuracy in malware detection compared with ML-based algorithms, confirming the effectiveness of the automatic feature learning characteristic of DL compared with traditional feature engineering methods.

Meidan et al. [237] implemented ML algorithms for precise IoT device identification by utilising network traffic features, which are then fed into a multi-stage classifier. The classifier categorises the devices that are connected to the network as IoT or non-IoT devices; the ML algorithms identify unauthorised links of IoT devices automatically and accordingly alleviate the disruptions that may occur due to threats.

In a previous study [238], the abnormal behaviour of IoT objects was profiled, and the generated dataset from profiling was used to train the classifier to detect abnormal behaviour. The author investigated how a partial variation (assuming that the attacker can utilise such changes for malicious purposes) of sensed data can influence the accuracy of the learning algorithm and used SVM and k-means clustering as experimental cases for examining the impact of such changes on the detection accuracy of both ML algorithms. The results showed that both algorithms (i.e. SVM and k-means) suffered from detection accuracy drops. The zero-day attacks are mostly variations of existing attacks; thus, the accuracy of the classifier in detecting variations and changes in the dataset is research topic for future investigation.

A system called 'IoT SENTINEL', which is based on the RF classification algorithm, was proposed in [239] to recognise the types of devices connected to an IoT system automatically and execute an action to restrain any of vulnerable connections accordingly to reduce damage that may be caused by compromised devices.

A previous study [240] developed an IDS for IoT by combining fuzzy c-means clustering [241] and the feature selection method PCA [177]. The results of the study indicated that the proposed method can increase detection effectiveness.

In [242], the authors proposed a framework to recognise all potential attack paths and alleviate the effects of attacks on the IoT system; the proposed framework contains a graphical security model. The framework consists of five connected stages starting with data processing, in which the information from the system and the security metrics is fed and processed. In the second stage, which is the security model generation, a gap model is generated; this model contains all potential attack paths in the IoT system; an attack path identifies the structure of the nodes that the intruder can compromise to gain access to the required node. In the third and fourth stages, the IoT network, including the attack paths, is visualised (i.e. security visualisation) and analysed (i.e. security analysis), respectively. Finally, the security model is updated on the basis of the analysis of the attack paths and patterns captured in the previous stages. However, this study used basic statistical analysis to obtain the security model; therefore, whether the proposed framework can be improved by integrating it with intelligent methods, such as ML or DL methods, should be investigated.

In [243], a solution was proposed to detect and restrain malware diffusion in an IoT network. The solution is based on fog computing, which can simultaneously maximise malware detection and minimise the possibility of privacy breach. The proposed malware detection system was constructed using an IDS, and deployment was accomplished at cloud and fog computing to avoid the restrictions on IDS deployment in smart objects [243]. The authors also presented a framework to show the possible application of malware dissemination restraint in IoT networks.

*C. Application layer*

Currently, most IoT services have application and user interfaces; for example, the Android platform is becoming a vital element for enabling the IoT system [49]. In the related security literature, a previous study [244]. showed the effective performance of DL in accurately detecting Android malware, and the authors of the study constructed a DL model to learn features from Android apps. Subsequently, the learning model was used to identify unspecified Android malware; the authors showed the effectiveness of using DL in Android malware detection in terms of performance accuracy and time efficiency, indicating that DL can be adapted to real-world applications.

A past work [188] proposed an Android malware detection method that utilises a CNN. With the application of the CNN, the significant features related to malware detection are learnt automatically from the raw data, thereby eliminating the need for manual feature engineering. The main advantage of using DL algorithms, such as CNNs, is that the network is trained to learn suitable features and execute classification conjointly, eliminating the extraction process required in traditional ML and consequently providing an end-to-end model [188].

The study in [206] proposed an architecture for securing the cyberspace of IoT systems, and the proposed architecture involves training ML algorithms to classify the system behaviour as normal or abnormal. They used GAN algorithms, which were integrated into the proposed architecture for preliminary study, whose results showed the effectiveness of DL-based architecture in detecting abnormal system behaviour.

Cybersecurity remains to be a serious challenge, especially with the steadily increasing number of objects connecting to the cyberspace, such as IoT. Cyberattacks, including zero-day attacks, are incessantly evolving; consequently, the vulnerabilities and opportunities open to attackers increase with the rapid growth of IoT. Many of these attacks are minor variations of formerly identified cyberattacks [245]. Therefore, the recent improvements in effective learning algorithms are significant. Effective learning algorithms can be trained to adapt to attack variations in the cyberspace with high-level feature abstraction capability; thus, they can provide resilient solutions to the variations of formerly identified cyberattacks or new attacks [245]. A previous study [245] proposed a DL model to enhance cybersecurity and enable attack detection in IoT systems and verified the appropriateness of the DL model in securing the cyberspace of IoT systems. Similarly, [246] proposed a distributed DL model to deliver accurate protection against cyberattacks and threats in fog-to-things computing and used an SAE algorithm to construct their learning model. The authors confirmed that the DL models are more suitable for such cyberattack protection than are traditional methods in terms of scalability, accuracy and false alarm rate.

Table 3 shows the comparison and summary of studies on ML and DL for IoT security.

*D. Enabling technology for ML/DL deployment for IoT security*

On the one hand, realising ML/DL to construct an intelligence-based security for IoT systems can be practically challenging because robust software and hardware requirements are required to implement such complex algorithms. On the other hand, the recent advancements in computational capability of tiny devices and in several ML/DL implementation platforms can result in successfully implementing these methods onboard devices, such as smartphones [247], or in fog and edge computing platforms [247]. As shown in Figure 10 the technology tools that can essentially enable ML/DL deployment for IoT security can be generally listed as a large growth of IoT data, robust software frameworks for facilitating the development of security model based on ML/DL methods and sophisticated hardware equipment to deploy the developed security model.

*Large growth of IoT data*: The large growth will result in producing large-volume data. The data contains useful information about the system behaviour under different modes, that is, 'normal and attack modes'. The current volume of data is considered larger than that in the past. Data are the main elements for successful implementing ML/DL-based systems. Therefore, additional data about system behaviour, which can be used to successfully enable ML/DL deployment for IoT security, are produced with the continued growth of IoT systems. However, several challenges related to creating security data remain. These challenges are discussed in the challenges and future direction section.

*Robust software frameworks:* A recent development of ML/DL has resulted in introducing several dedicated ML/DL implementation frameworks and libraries that can empower ML/DL deployment for IoT security that can enable ML/DL deployment for IoT security. The learning algorithms are continuously developing. However, building and deploying them successfully can be challenging without effective frameworks and dedicated libraries. Currently, several frameworks allow the building of models that can offer an enhanced level of abstraction with minimal programming

| Study | Table 3 Comparison and summary of studies on ML and DL for IoT security ||||||||||||||||| Attack surfaces secured ||| Threats detected or security application |
|---|---|---|---|---|---|---|---|---|---|---|---|---|---|---|---|---|---|---|---|---|---|---|
| | ML |||||||||  DL |||||||| | | | | |
| | Supervised approaches ||||||| Unsupervised approaches || Supervised approaches ||| Unsupervised approaches |||| Hybrid approaches | | | | | |
| | DTs | SVMs | NB | KNN | RF | AR algorithms | EL | k-means clustering | PCA | ANNs | CNNs | RNNs | AEs | RBMs | DBNs | GANs | EDLNs | RL | Perception Layer | Network layer | Application Layer | |
| [224] | – | – | – | – | – | – | – | – | – | ✓ | – | – | – | – | – | – | – | – | – | ✓ | – | Routing attack detection |
| [215] | – | – | – | – | – | – | – | – | – | – | – | – | – | – | – | – | – | ✓ | ✓ | – | – | Jamming attacks |
| [128] | – | ✓ | – | ✓ | – | – | ✓ | – | – | – | – | – | – | – | – | – | – | – | – | – | ✓ | False data injection attacks |
| [225] | – | ✓ | – | – | – | – | – | – | – | – | – | – | – | – | – | – | – | – | – | ✓ | – | Intrusion detection |
| [197] | – | – | – | – | – | – | – | – | – | – | ✓ | – | – | – | – | – | – | – | – | ✓ | – | Malicious behaviour detection |
| [228] | – | – | – | – | – | – | – | – | – | ✓ | – | – | – | – | – | – | – | – | – | ✓ | – | Data tampering |
| [216] | – | – | – | – | – | – | – | – | – | ✓ | – | – | – | – | – | – | – | – | ✓ | – | – | Spoofing attack detection |
| [212] | – | – | – | – | – | – | – | – | – | – | – | – | – | – | – | – | – | ✓ | ✓ | – | – | Jamming attacks |
| [199] | – | – | – | – | – | – | – | – | – | – | – | – | ✓ | – | – | – | – | – | – | ✓ | – | Malware detection |
| [233] | – | – | – | – | – | – | – | – | – | ✓ | – | – | ✓ | – | – | – | – | – | – | ✓ | – | Impersonation attacks |
| [235] | – | – | – | ✓ | – | – | – | – | – | – | – | – | ✓ | – | – | – | – | – | – | ✓ | – | Impersonation attacks |
| [245] | – | – | – | – | – | – | – | – | – | ✓ | – | – | – | – | – | – | – | – | – | – | ✓ | Cyberattacks |
| [246] | – | – | – | – | – | – | – | – | – | – | – | – | ✓ | – | – | – | – | – | – | – | ✓ | Cyberattacks |
| [202] | – | – | – | – | – | – | – | – | – | – | – | – | – | ✓ | – | – | – | – | – | ✓ | – | Network anomaly detection |
| [236] | – | – | – | – | – | – | – | – | – | ✓ | – | – | – | – | – | – | – | – | – | ✓ | – | DDoS attack detection |
| [204] | – | – | – | – | – | – | – | – | – | – | – | – | – | – | ✓ | – | – | – | – | ✓ | – | Malicious attack detection |
| [217] | – | ✓ | – | – | – | – | – | – | – | ✓ | – | – | – | – | – | – | – | – | ✓ | – | – | Authentication |
| [188] | – | – | – | – | – | – | – | – | – | – | ✓ | – | – | – | – | – | – | – | – | – | ✓ | Malware detection |
| [237] | – | – | – | – | – | – | ✓ | – | – | – | – | – | – | – | – | – | – | – | – | ✓ | – | IoT device identification (authorisation) |
| [238] | – | ✓ | – | – | – | – | – | ✓ | – | – | – | – | – | – | – | – | – | – | – | ✓ | – | Data tampering and abnormal behaviour |
| [239] | – | – | – | – | ✓ | – | – | – | – | – | – | – | – | – | – | – | – | – | – | ✓ | – | Authorisation |
| [240] | – | – | – | – | – | – | – | – | ✓ | – | – | – | – | – | – | – | – | – | – | ✓ | – | Intrusion detection |
| [206] | – | – | – | – | – | – | – | – | – | – | – | – | – | – | ✓ | – | – | – | – | ✓ | – | Abnormal behaviour detection |

complications. These ML/DL frameworks and dedicated libraries support several programming languages and are built with a GPU to optimise the training process of DL algorithms. [248]. Consequently, these libraries can enable the ML/DL deployment for IoT security by providing an efficient and easy implementation of ML/DL. Therefore, researchers and scientists on security requirements mostly focus on applying and optimising such algorithms rather than building them from scratch which can be time-consuming and entail high costs. These libraries include TensorFlow [249], convolutional architecture for fast feature embedding [250], Theano [251], deeplearning4j [252], Torch [253], Neon and MXNet [254] (for additional DL Libraries, see [252]).

*Effective deployment strategies:* The ML/DL model for IoT security can be deployed on-board, on cloud computing or on edge computing. The search for the optimal deployment strategy is vital for the ML/DL-based model implementation in real life to secure IoT devices and systems. The deployment of ML/DL for IoT system security must consider the limited resource of IoT devices (i.e. limited computational power and memory size), real-time threats detection and response and accessibility to frequent updating of security models (updating the model to detect newly emerging threats).

For the limited resource of IoT devices, offloading the ML/DL execution can be a practical solution in terms of computational power, memory size and performance and a suitable means for regular updates of the ML/DL model. However, offloading may lead to high latency [255] which may not satisfy the required real-time detection systems for practical IoT systems. The second choice for deploying ML/DL for IoT system security can be accomplished by deploying the ML/DL security model on the IoT device which is not concerned with communication quality [255]. However, deploying ML/DL on the IoT device remain challenging. The main feature of IoT devices is limited, thereby indicating limited computational power and memory size. However, applying ML/DL-based models involves computational power, high power consumption and sizable memory to store the model. Moreover, current ML/DL frameworks are commonly built on third-party libraries, which renders the migration to onboard deployment extra challenging [255].

On the one hand, the abovementioned solution can still be optimised in the future to successfully satisfy the requirements for deploying ML/DL for IoT system security. On the other hand, edge computing can bridge the gap between the deployment of cloud computing with powerful resources but high latency and the onboard deployment with no latency but limited resources. The deployment of ML/DL for IoT system security using edge computing can enable effective computation power over onboard deployment that is executed at the edge of the network. Therefore, the processing is performed near the data sources and has less latency than the deployment on the cloud. Edge computing can process downstream data from cloud services to IoT devices and upstream data from IoT devices to cloud services (e.g. smartphone that operates between body sensors and the cloud) [256]. By implementing an edge computing framework, the ML/DL security model for the IoT system can be placed near the network edge, operate as guards to detect malicious behaviour, secure devices and reduce the severity of eavesdropping threats by approaching the IoT devices [257].

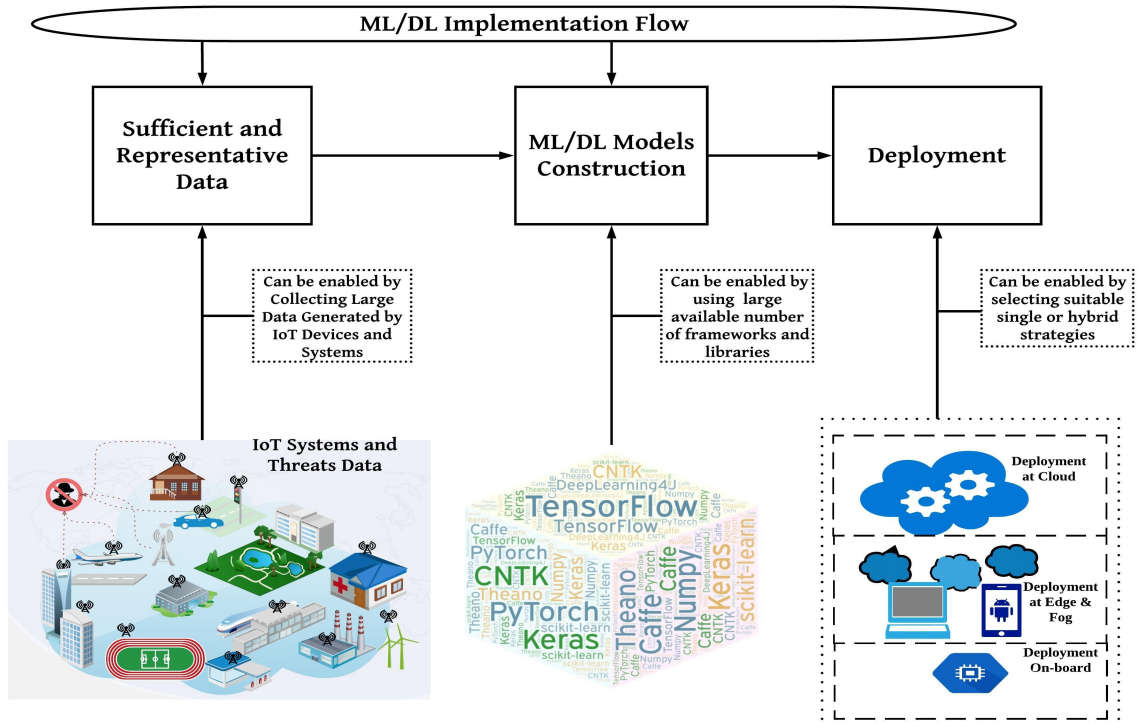

*Figure 10 Illustration the technology tools that can essentially enable ML/DL deployment for IoT security*

## VI. Issues, challenges and future directions

In this section, we present a list of Issues, challenges and future directions for using ML and DL methods to mitigate security weakness IoT systems, which are classified based on data, learning strategies, IoT environments, inherent ML and DL Challenges, opportunities to integrated ML/DL with other technology, computational complexity issues and security vs other trades off requirements.

### A. IoT data related issues

#### 1) Availability of security related datasets

The general purpose of learning algorithms is capturing the patterns from the available partial training dataset and then constructing a model to categorise the new inputs on the basis of the learnt patterns. In this process, a question to investigate is the volume of training data required to train the learning algorithms sufficiently for these algorithms to be generalised for new input in the given domain [258]. In the context of the application of ML and DL for IoT security, the major challenge encountered by ML and DL, in general, and the supervised ML and DL methods in particular, is how to extract or generate a realistic and high-quality training dataset that contains various possible attack types. A high-quality training dataset is an essential ingredient to train the ML and DL algorithms accurately. The training datasets should be comprehensive and diverse. They should contain information that reflects nearly all of the strategies of real-world attacks because these training datasets are the basis for obtaining model knowledge. This condition can directly influence model accuracy. Given that IoT systems generate large volumes of data, real-time data streaming data quality maintenance remains a challenge.

A vital future research direction is the use of crowd-sourcing methods for generating datasets related to IoT threats and attacks. Rich datasets that include nearly all attack patterns should be generated for training ML and DL algorithms. Furthermore, such datasets can be used to benchmark the accuracy of newly proposed algorithms against that of existing methods for attack detection. Although generating collaborative IoT threat datasets, which can be continuously updated with new attacks, is of great importance, it is challenging technically due to the wide diversity of IoT devices. Furthermore, a privacy issue prevails because datasets may contain sensitive or critical information that are not meant to be shared publicly, specifically for industrial and medical IoT devices.

#### 2) Learning to secure IoT with low-quality data

Most of the proposed DL representations are generally for high-quality data [195]. However, IoT systems comprise heterogeneous connected devices, and large-scale streaming, leading to the possibility of high-noise and corrupted data to be gathered from such systems[198, 259]. Therefore, learning to secure IoT systems requires effective DL models that can handle and learn from low-quality data, particularly when obtaining high-quality training data is practically infeasible. Therefore, multi-modal and effective DL models should be developed to secure IoT systems with large-scale streaming, heterogeneous and high-noise data.

#### 3) Augmentation of IoT security data to improve learning algorithm performance

Intuitively, richer the data that ML and DL algorithms have to learn from, the more accurate they can be [260]. Although obtaining a large dataset is relatively easy in certain domains, such as image and natural language processing, acquiring a large dataset for ML and DL is relatively difficult in the domain of data security in IoT systems. Therefore, finding alternative means to obtain substantial amounts of data in this domain is desirable. Data augmentation is used to expand limited data by generating new samples from existing ones. In the augmentation of IoT security data, the limited amount of existing IoT security samples can be utilised to generate new samples.

The key challenge in data augmentation is producing new data samples that preserve the appropriate data distribution for each class, normally necessitating domain knowledge [192, 261]. In view of this problem, suitable methods for the augmentation of IoT security data should be investigated to improve the classification accuracy of learning methods.

### B. Learning Strategies for Effective IoT Security

#### 1) Zero-day attacks on IoT

The main advantage of ML and DL methods over traditional security methods, such as the threat signature-based method, is their capability to detect zero-day attacks. Zero-day attacks, which are evolving threats, were previously anonymous to detection systems. These attacks have varying potentials, such as metamorphic malware attacks that automatically reprogram themselves each time they are circulated or transmitted. Consequently, detecting these malware attacks by traditional methods is difficult [262, 263]. The number of emerging IoT security threats [24]., such as zero-day attacks, is continuously growing at an alarming rate [24]. For example, the Mirai botnet and its derivations are becoming an alarming threat to the security of IoT systems [7, 9]. The development of the recent derivation of the Mirai botnet, Satori, proves that other malicious IoT botnets are emerging to exploit known and zero-day vulnerabilities [264].

On the one hand, the recent derivations of Mirai suggest that IoT malware will continue growing because Mirai's open source code allows creators of IoT malware to produce new variants of Mirai that exploit known and zero-day vulnerabilities to attack IoT devices [7, 9, 24, 264]. On the other hand, having the ability to monitor and control IoT security intelligently provides an important solution to these new attacks or the zero-day attack and its variations. ML and DL algorithms are powerful analysis tools for learning normal or abnormal behaviour on the basis of the interactions among the IoT systems and devices within the IoT ecosystem. The input data from each element of an IoT system and its devices can be collected and examined to determine normal patterns of interaction and consequently identify malicious behaviour at an early stage. Moreover, in view of the capability of ML and DL methods to learn from existing samples to predict future unknown samples intelligently, these methods have the

4potential to predict new attacks, which, in many cases, are simple derivations and mutations of previous attacks. Therefore, IoT security systems need to advance from the simple facilitation of the secure communication between devices to intelligent security enabled by DL and ML methods.

*2) Lifelong Learning for learning IoT threats*

One of main characteristics of the IoT environment is dynamism; several new things join and numerous objects leave the system given the numerous and diverse IoT devices used to manage different applications and scenarios. Given IoT's nature, normal structures and patterns of IoT systems may considerably change with time, and threats and attacks targeting the IoT system may likewise persistently vary with time. Therefore, distinguishing between normal and abnormal IoT system behaviour cannot be always pre-defined. Thus, the frequent updating of security models is required to handle and understand IoT modifications. In an actual IoT environment, the short-term learning of threats and attacks targeting IoT systems may be ineffective for long-term protection. Consequently, the concept of lifelong learning can hold realistic significance in long-term real-world applications. The concept of lifelong machine learning [106, 265, 266] is directed towards the construction of a model that can perform the retraining process repeatedly for the learning of new emerging patterns related to each behaviour. The model should be able to adapt to and learn from new environments continuously [106, 265, 266]. Researchers have reported that the further they trained the algorithm with the latest features of known DDoS attacks, the more they improved the detection probabilities for known and unknown DDoS attacks. The ANN algorithm learns from training samples and then detects zero-day attack features, which are comparable to the features on which it was trained [236]. Therefore, frequently updating the training samples is important for developing effective real-world security models for IoT-related threats.

*3) Transfer learning*

Transfer learning refers to the idea of transferring knowledge from a domain with sufficient training data to a domain with insufficient training data. The main purpose of transfer learning is to reduce the time and effort required for the new learning process. The main concern in transfer learning deals with the part of knowledge that can be transferred as knowledge that is common between the domains. Therefore, transferring such knowledge is useful. Meanwhile, transferring knowledge that is specific for a particular domain and does not hold any importance to other domains must be avoided [267].

The concept of transfer learning can be useful for securing IoT systems, which comprise different elements, such as devices, WSNs and cloud computing. The security of these elements has already been extensively studied, and well-established training samples on different attacks have been generated. Consequently, if transfer learning is accomplished successfully from the elements of the IoT, then such learning may significantly improve the security performance of the entire IoT system with less effort and cost in constructing training samples.

*C. ML and DL for IoT security in interdependent, interconnected and interactive environments*

In this section, we present the opportunities for using ML and DL methods to mitigate internal security issues arising from the structures of IoT systems, which are interdependent, interconnected and interactive environments.

As explained previously, with the rapid increase in the number of IoT devices, the collaboration among devices is becoming increasingly autonomous; i.e. they require reduced human involvement. IoT devices no longer simply interact with one another like devices within a network. Many current IoT devices are designed to achieve the vision of a smart city, in which many of the devices are controlled by other devices or depend on the operational condition of other devices or the surrounding environment

The advantage of using ML and DL in securing IoT devices in such an environment is that these methods can be developed to go beyond simply understanding the operational behaviour of specific devices to understanding the operational behaviour of entire systems and their devices.

Moreover, IoT systems connect billions of devices; thus, not only the surface of the attack but also the magnitude of the attack should be considered in IoT systems. With these densely interconnected devices, an infected thing can result in a destructive attack that infects a considerable number of things at a large scale, even affecting a substantial part of a city.

For interconnected systems, the benefit of using ML and DL for securing IoT devices is that ML the DL methods can provide intelligence to systems for detecting abnormal behaviours of a thing or groups of things and thus automatically respond at an early stage. This strategy may mitigate the impact of the attack and lead to learning for the prevention of future occurrences of similar attacks on the basis of a solid understanding of the current causes.

Along the same direction, ML and DL can be effective for securing IoT devices in an interactive environment In SIoT, suitable instructions should be established for objects to choose their appropriate friends as these impacts the service outputs built on top of the social network [104]. The advancement in SIoT increases critical security and privacy concerns regarding the disclosure of sensitive information related to the objects [105].

ML and DL methods can potentially contribute to securing the integration of social networking into IoT. However, this direction is still at its infancy and needs further investigation.

*D. ML and DL Challenges*

*1) Possible misuse of ML and DL algorithms by attackers (breaking cryptographic implementations by ML and DL methods)*

Recent advances in ML and DL algorithms have enabled them to be used in breaking cryptographic implementations. For example, two previous studies [129, 130] used ML to break cryptographic systems using SVMs, which outperformed the template attack. Similarly, the authors in [189] investigated different DL algorithms to break cryptographic systems and reported that DL can break cryptographic systems. Specifically,

CNN and AE algorithms performed better than did ML algorithms (SVM and RF) and the rational profiling method template attack.

A previous study showed that RNNs can learn decryption. Specifically, an RNN with a 3000-unit LSTM can learn the Enigma machine decryption function by learning effective internal representations of these ciphers; the results suggested that DL algorithms, such as RNNs, can capture and learn the algorithmic representations of polyalphabetic ciphers for cryptanalysis [268].

*2) Privacy of ML and DL*

Recent studies [269-271] have shown that ML and DL algorithms can leak data. Privacy-preserving ML and DL algorithms are vulnerable to dominant attacks [269]. A study showed that federated, distributed or even decentralised DL methods are easily broken and unable to maintain training set privately [269]. The authors developed an attack to manipulate the real-time nature of a learning process in which the adversary was allowed to train a GAN that creates samples similar to those in the targeted training dataset, which was supposed to be private; the samples produced by the GAN were supposed to originate from the same distribution as the training dataset [269]. Therefore, DL algorithms themselves are vulnerable to potential attacks when generating the training data. Consequently, the attackers can build a DL system that can recognise how DL-based detection methods work and thus generate attacks which cannot be detected easily. This area of research is still at its infancy and needs to be investigated further to find the appropriate solution to such an issue.

*3) Security of ML and DL methods*

Researchers have recently investigated various threats that can be launched against ML and DL algorithms. These algorithms are susceptible to many threats that either decrease the accuracy and performance of the classifiers or expose sensitive data used in the training process of the classifiers. Examples of the potential threats that can be utilised by attackers include poisoning, evasion, impersonation and inversion attacks s [272]. Poisoning is a threat in which the attacker injects malicious samples with incorrect labels into the training dataset to modify training data distribution, decrease the discrimination power of the classifier in distinguishing between the normal and abnormal behaviour of the system, and ultimately decrease classifier accuracy and performance. Such attacks can be potentially launched against ML algorithms that need to dynamically update their training sets and learning models to adapt to the new attacks features, such as ML algorithms for malware detection [272, 273]. The second possible attack on ML and DL is the evasion attack. This attack is based on generating adversarial samples by modifying the attack features to be slightly different from the malicious samples used to train the model; consequently, the probability of the attack being detected by the classifier is decreased, and the attack avoids detection, thereby reducing the performance of the system remarkably [272]. The third possible attack is impersonation. In this attack, the attacker attempts to mimic the data samples to deceive the ML algorithms to classifying the original samples with different labels incorrectly from the impersonated ones [272, 274, 275]. The last possible attack is inversion, which exploits the application program interfaces presented to the users by the current ML platform to collect roughly the necessary information about the pre-trained ML models [271, 276]. Subsequently, this extracted information is used to perform reverse engineering to obtain the sensitive data of users. This kind of attack violates the privacy of users by exploring the data, which are sensitive in certain cases (e.g. patients' medical data), inserted in the ML models [277, 278].

*4) Insights into DL architecture*

ML and DL methods change the means through which a computer solves a problem, from instructing the computer what to do programmatically to training the computer what to do intelligently (learning from experience). However, despite the progress achieved by DL algorithms in many applications, a theory that can describe why and how DNNs run depending on their architecture has not yet been established. Such a theory can be significant in comprehending the quantity of data or the number of the layers required to achieve the desired performance. The theory can also facilitate the reduction of the resources (e.g. time, energy and memory) required to construct a DL model [180], thereby providing a sophisticated but lightweight DL model that is useful for resource-constrained systems, such as IoT devices. Establishing a lightweight DL model is a significant step towards the implementation of on-board security systems for IoT devices. Thus, this topic needs further exploration in future studies.

*E. Integrating DL/ML with Other Technology for IoT Security*

*1) Implementation of ML and DL at the edge*

Edge computing has become an essential technology in providing IoT services. Edge computing immigrates service provision from the cloud to the network edge, which holds a potential solution in the IoT era [179, 279]. Implementing DL and ML at the edge for IoT security can minimise delays, realise near-real-time detection systems, improve energy efficiency and enhance the scalability of lightweight IoT objects. Such implementation can offer an effective framework for data processing with reduced network traffic load. However, edge computing is still at its infancy, and its implementation is accompanied by several challenges. Further research needs to be conducted to explore and develop effective strategies for implementing DL and ML at the edge to provide real-time IoT security.

*2) Synergic integration of ML and DL with blockchain for IoT security*

Blockchain is an emerging technology that uses cryptography to secure transactions within a network. A blockchain delivers a decentralised database (called 'digital ledger') of transactions, of which each node on the network is aware [280]. The network is a chain of devices (e.g. computers) that all need to endorse a transaction before it can be verified and recorded [280]. In other words, a blockchain is simply a data structure that allows the production and distribution of a 'tamper-proof digital ledger' of exchanges [281]. The decentralised architecture of a blockchain is antithetical to the security issues that are inherent in a centralised architecture. Using decentralised database architecture, transaction authentication depends on the approval of many parties in systems rather than of a single authority, as is common practice in centralised systems. Therefore, blockchain systems can render transactions relatively more secure and transparent than those in centralised systems. IoT systems are distributed by nature. Thus, the distributed digital ledger blockchain can play a significant role in securing IoT systems.

ML and DL are concerned with training machines to learn from real-world samples to act autonomously and intelligently. The goal of ML and DL methods is to allow the machines to become smart machines. The simplified definitions of both technologies (i.e. ML/DL and blockchain) reveal that a synergic relation can be obtained by combining both technologies to achieve a fully functional IoT security system. Firstly, ML and DL may assist blockchain technology in realising smart decision making and improved evaluation, filtering and comprehension of data and devices within a network to facilitate the effective implementation of blockchain for enhanced trust and security services for IoT systems. Secondly, blockchain may assist ML and DL by providing a large volume of data because blockchain is a decentralised database that stresses the importance of data distribution among several nodes on a specific network. The availability of big data is a main factor in establishing an accurate ML- and DL-based model. Therefore, with the increase in the volume of data to be analysed, particularly security-related data, the accuracy of ML and DL methods can be considerably increased and generalised to develop a security model with enhanced reliability.

*F. Computational complexity*

IoT devices are resource-constrained devices. The resources of IoT devices (things), such as memory, computation and energy, which are required for ML and DL deployment, are limited and create a crucial bottleneck in the adoption of DL and ML for real-time on-board implementation [282]. The current solutions of computation offloading and execution in the cloud suffer from high wireless energy overhead. Moreover, the availability of the applications for such solutions is based on the network conditions. Consequently, if the network connectivity is weak, then cloud offloading will be unattainable, leading to the unavailability of the applications. Another recent solution which may advance the implementation of ML and DL for IoT security is the development of edge computing GPUs (mobile GPU). However, GPUs on mobile can still consume considerable mobile battery reserves [282].

On the one hand, enhancing GPU-based solutions and proposing an efficient offloading strategy are important in advancing the implementation of ML- and DL-based IoT security to enhance the performance of IoT DL applications in IoT systems with cloud and edge computing [179]. On the other hand, ML and DL frameworks that can efficiently reduce computational complexity should be developed. Developing real-time detection and protection systems are important for providing effective security mechanisms, particularly for large-scale IoT systems. Thus, reducing computational complexity holds practical importance in future research.

*G. Security vs Trade-offs in IoT Applications*

The existing security trade-offs, such as that between availability and safety, are another challenge to the achievement of a robust security scheme for IoT systems. Moreover, the importance of various security trade-offs differs from one IoT application to another. For example, an IoMT system should provide a security scheme, but it should also offer the flexibility of being accessible in emergency situations. When a patient with an implanted IoMT, which monitors his or her health conditions, is suddenly in an emergency situation, easy access to the IoMT device is the first priority in saving his or her life. Therefore, creating a design that balances providing a robust security scheme to protect the implanted IoMT and guaranteeing accessibility of such devices during emergency situations is necessary. Such a trade-off between security and safety poses a critical challenge. An appropriate balance between patient safety and device security is an important parameter to be considered in the design phase [47, 62]. ML and DL methods mainly aim to provide intelligence and contextual awareness to devices; therefore, these methods can better mitigate security trade-off issues than can traditional access control methods.

Similarly, other applications of IoT have different security trade-offs in accordance with the diverse implemented environments. Given the required security level and security trade-offs in specified IoT applications, security design should satisfy different operation modes within the given applications. Future research may utilise the intelligence capability of ML and DL methods to design security schemes that can effectively satisfy various security trade-offs under different operation modes within a specified application.

## VII. CONCLUSION

The requirements for securing IoT devices have become complex because several technologies, from physical devices and wireless transmission to mobile and cloud architectures, need to be secured and combined with other technologies. The advancement in ML and DL has allowed for the development of various powerful analytical methods that can be used to enhance IoT security.

In this survey, various IoT security threats and IoT attack surfaces are discussed. A comprehensive review of the potential uses of ML and DL methods in IoT security is provided. These methods are then compared at the end of each subsection in terms of their advantages, disadvantages and applications in IoT security. Afterward, the uses of the ML and DL methods for securing the main IoT layers (i.e. perception, network and application layers) are reviewed. Finally, an extensive list of issues, challenges and future directions related to the use of ML and DL in effectively securing IoT systems are presented and classified according to data; learning strategies; ML and DL for IoT security in the interdependent, interconnected and interactive environments of IoT systems; diverse security trade-offs in IoT applications and synergic integration of ML and DL with blockchain for IoT security.

This survey aims to provide a useful manual that can encourage researchers to advance the security of IoT systems from simply enabling secure communication among IoT components to developing intelligent end-to-end IoT security-based approaches.

**Acknowledgements**: This work is supported by NPRP grant #8-408-2-172 from the Qatar National Research Fund (a member of Qatar Foundation). The statements made herein are solely the responsibility of the authors.

*List of Acronyms*

| Acronym | Description |
|---|---|
| 6LoWPAN | Combination IPv6and Low-power Wireless Personal Area Networks |
| AEs | Auto-encoders |
| ANN | Artificial Neural Network |
| CNN | Convolutional Neural Network |
| CCTV | Closed-Circuit Television |
| CPS | Cyber-physical System |
| ARs | Association Rules |
| DL | Deep Learning |
| DBN | Deep Belief Network |
| DNN | Deep Neural Network |
| DoS | Denial of Service |
| DDoS | Distributive Denial of Service |
| DRL | Deep Reinforcement Learning |
| DT | Decision Tree |
| EL | Ensemble Learning |
| EDLNs | Ensemble Deep Learning Networks |
| GAN | Generative Adversarial Network |
| GPS | Global Positioning System |
| GPU | Graphics Processing Unit |
| IoT | Internet of Things |
| IoMT | Internet of Medical Things |
| KNN | K-nearest neighbour |
| LAN | Local Network Area |
| LSTM | Long Short-term Memory |
| MitM | (Man-in-the-Middle |
| NB | Naive Bayes |
| NFC | Near Field Communication |
| PCA | Principal Component Analysis |
| RBMs | Restricted Boltzmann Machines |
| ReLU | Rectified Linear Units |
| RNN | Recurrent Neural Network |
| RF | Random-Forest |
| RFID | Radio Frequency Identification |
| (SIoT) | Social Internet of Things |
| (SQL) | structured query language |
| SNs | Sensor networks |
| SVMs | Support Vector Machines |
| UWB | ultra-wide bandwidth |
| WAN | Worldwide Network Area |
| WSN | Wireless Sensor Network |
| Wi-Fi | Wireless Fidelity |

35

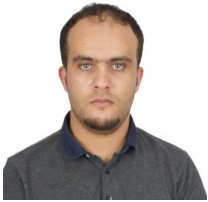
**Mohammed Ali Al-Garadi** received his PhD in Computer science from University of Malaya, Malaysia in 2017. He obtained his B. Tech and M.Tech. degree in electronic and communication engineering from Jawaharlal Nehru Technological University, Hyderabad, India. Currently, he is a researcher in Qatar University in joint collaborative project between Qatar University, University of Idaho, USA, and Temple University, USA. He has published several research articles in refereed journals and conferences. He served as reviewer for several journals including, IEEE Communications Magazine, IEEE Transactions on Knowledge and Data Engineering, IEEE Access, Future Generation Computer Systems, Computers & Electrical Engineering, and Journal of Network and Computer Applications. He also received several national and international awards during his PhD research. His research interests include Big data analytics, Machine learning, Deep learning, Cybersecurity, and IoT systems.

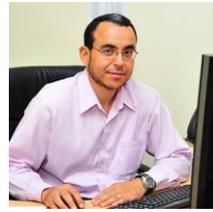
**Amr Mohamed** (S'00–M'06–SM'14) received the M.S. and Ph.D. degrees in electrical and computer engineering from The University of British Columbia, Vancouver, Canada, in 2001 and 2006, respectively. He was an Advisory IT Specialist with the IBM Innovation Centre, Vancouver, from 1998 to 2007, taking a leadership role in systems development for vertical industries. He has over 25 years of experience in wireless networking research and industrial systems development. He is currently an Associate Professor with the College of Engineering, Qatar University, and the Director of the Cisco Regional Academy. His research interests include wireless networking and edge computing for IoT applications. He has authored or co-authored over 150 refereed journal and conference papers, textbook, and book chapters in reputed international journals and conferences. He holds three awards from IBM Canada for his achievements and leadership, and three best paper awards, latest from IEEE/IFIP NTMS 2015 in Paris. He has served as a Technical Program Committee (TPC) Co-Chair for workshops in IEEE WCNC 2016 and a Co-Chair for the technical symposia of international conferences, including GLOBECOM 2016, Crowncom 2015, AICCSA 2014, IEEE WLN 2011, and IEEE ICT 2010. He has served on the organization committee of many other international conferences as a TPC Member, including the IEEE ICC, GLOBECOM, WCNC, LCN, and PIMRC, and a technical reviewer for many international IEEE, ACM, Elsevier, Springer, and Wiley journals. He is serving as a Technical Editor for the Journal of Internet Technology and the International Journal of Sensor Networks.

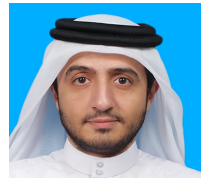
**Abdulla Khalid Al-Ali**, Ph.D. obtained his Master degree in Software Design Engineering and PhD degree in Computer Engineering from Northeastern University in Boston, MA, USA in 2008 and 2014, respectively. He is an active researcher in Cognitive Radios for smart cities and vehicular ad-hoc networks (VANETs). He has published a number of peer-reviewed papers in journals and conferences. Dr. Abdulla is currently head of the Technology Innovation and Engineering Education (TIEE) at the College of Engineering in Qatar University.

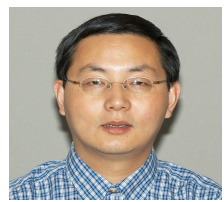
**Xiaojiang (James) Du** is a professor in the Department of Computer and Information Sciences at Temple University, Philadelphia, USA. Dr. Du received his B.S. and M.S. degree in electrical engineering from Tsinghua University, Beijing, China in 1996 and 1998, respectively. He received his M.S. and Ph.D. degree in electrical engineering from the University of Maryland College Park in 2002 and 2003, respectively. His research interests are security, wireless networks, and systems. He has authored over 230 journal and conference papers in these areas, as well as a book published by Springer. Dr. Du has been awarded more than 5 million US dollars research grants from the US National Science Foundation (NSF), Army Research Office, Air Force, NASA, the State of Pennsylvania, and Amazon. He won the best paper award at IEEE GLOBECOM 2014 and the best poster runner-up award at the ACM MobiHoc 2014. He serves on the editorial boards of three international journals. Dr. Du served as the lead Chair of the Communication and Information Security Symposium of the IEEE International Communication Conference (ICC) 2015, and a Co-Chair of Mobile and Wireless Networks Track of IEEE Wireless Communications and Networking Conference (WCNC) 2015. He is (was) a Technical Program Committee (TPC) member of several premier ACM/IEEE conferences such as INFOCOM (2007 - 2018), IM, NOMS, ICC, GLOBECOM, WCNC, BroadNet, and IPCCC. Dr. Du is a Senior Member of IEEE and a Life Member of ACM.

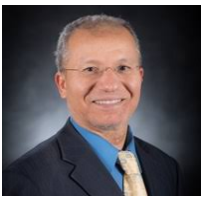

**Mohsen Guizani** received the B.S. (with distinction) and M.S. degrees in electrical engineering, the M.S. and Ph.D. degrees in computer engineering from Syracuse University, Syracuse, NY, USA, in 1984, 1986,1987, and 1990, respectively. He is currently a Professor and the ECE Department Chair at the University of Idaho, USA. Previously, he served as the Associate Vice President of Graduate Studies, Qatar University, Chair of the Computer Science Department, Western Michigan University, and Chair of the Computer Science Department, University of West Florida. He also served in academic positions at the University of Missouri-Kansas City, University of Colorado-Boulder, Syracuse University, and Kuwait University. His research interests include wireless communications and mobile computing, computer networks, mobile cloud computing, security, and smart grid. He currently serves on the editorial boards of several international technical journals and the Founder and the Editor-in-Chief of Wireless Communications and Mobile Computing journal (Wiley). He is the author of nine books and more than 500 publications in refereed journals and conferences. He guest edited a number of special issues in IEEE journals and magazines. He also served as a member, Chair, and General Chair of a number of international conferences. He received the teaching award multiple times from different institutions as well as the best Research Award from three institutions. He was the Chair of the IEEE Communications Society Wireless Technical Committee and the Chair of the TAOS Technical Committee. He served as the IEEE Computer Society Distinguished Speaker from 2003 to 2005. He is a Fellow of IEEE and a Senior Member of ACM.